\documentclass[twocolumn]{aastex631}

\usepackage{amsmath, amsfonts}
\usepackage{amssymb,amsmath,verbatim,mathtools,
needspace,enumitem,etoolbox,graphicx,physics,
microtype,afterpage,xspace,tabularx,lmodern,multirow,placeins}

\usepackage{xcolor}

\newcommand{\code}[1]{\texttt{#1}}

\newcommand{\nn}{\nonumber}
\newcommand{\be}{\begin{equation}}
\newcommand{\ee}{\end{equation}}
\def\be{\begin{equation}}
\def\ee{\end{equation}}

\usepackage{aas_macros}

\newcommand{\Msun}{\ensuremath{M_\odot}\xspace}

\newcommand{\Rsun}{\ensuremath{R_\odot}\xspace}
\newcommand{\kms}{\ensuremath{\rm km/s}\xspace}

\newcommand{\Spar}{\texttt{Par}\xspace}
\newcommand{\Sperp}{\texttt{Perp}\xspace}
\newcommand{\Siso}{\texttt{Iso}\xspace}
\newcommand{\Snat}{\texttt{Natal}\xspace}
\newcommand{\chieff}{\ensuremath{\chi_{\rm eff}}\xspace}

\newcommand{\thetaK}{\ensuremath{\theta_{k}}\xspace}
\newcommand{\phiK}{\ensuremath{\phi_{k}}\xspace}
\newcommand{\thetaS}{\ensuremath{\theta_{S}}\xspace}
\newcommand{\phiS}{\ensuremath{\phi_{S}}\xspace}
\newcommand{\thetaLL}{\ensuremath{\theta_{LL_i}}\xspace}
\newcommand{\thetaLS}{\ensuremath{\theta_{LS}}\xspace}

\newcommand{\subheading}[1]{\noindent\textbf{\textit{#1:}}}

\def\apjl{\rm{ApJ}}
\def\apjs{\rm{ApJS}}
\def\aap{\rm{A\&A}}

\allowdisplaybreaks

\begin{document}

\pagenumbering{arabic}

\title{Revising the Spin and Kick Connection in Isolated Binary Black Holes}

\author[0000-0002-2536-7752]{Vishal\,Baibhav}\footnote{NASA Einstein Fellow; \href{mailto:vb2630@columbia.edu}{vb2630@columbia.edu}}
\affiliation{Columbia Astrophysics Laboratory, Columbia University, 550 West 120th Street, New York, NY 10027, USA}
\affiliation{Center for Interdisciplinary Exploration and Research in Astrophysics (CIERA), Northwestern University, 1800 Sherman Avenue, Evanston, IL 60201, USA}

\author[0000-0001-9236-5469]{Vicky\,Kalogera}
\affiliation{Center for Interdisciplinary Exploration and Research in Astrophysics (CIERA), Northwestern University, 1800 Sherman Avenue, Evanston, IL 60201, USA}
\affiliation{Department of Physics and Astronomy, Northwestern University, 2145 Sheridan Road, Evanston, IL 60208, USA}
\affiliation{NSF-Simons AI Institute for the Sky (SkAI), 172 E. Chestnut Street, Chicago, IL 60611, USA}

\journalinfo{}

\begin{abstract}
The origin of black hole (BH) spins remains one of the least understood aspects of BHs. Despite many uncertainties, it is commonly assumed that if BHs originated from isolated massive star binaries, their spins should be aligned with the orbital angular momentum of the binary system. This assumption stems from the notion that BHs inherit their spins from their progenitor stars. In this study, we relax this long-held viewpoint and explore various mechanisms that can spin up BHs before or during their formation. In addition to natal spins, we discuss physical processes that can spin BHs isotropically, parallel to natal kicks, and perpendicular to natal kicks. These different mechanisms leave behind distinct imprints on the observable distributions of spin magnitudes, spin--orbit misalignments, and the effective inspiral spin of merging binaries. In particular, these mechanisms allow even the binaries originating in the field to exhibit precession and retrograde spin ($\chieff<0$). This broadens the parameter space allowed for isolated binary evolution into regimes which were previously thought to be exclusive to dynamically assembled binaries.
\end{abstract}

\section{Introduction}
\label{sec:intro}

\begin{nolinenumbers}
\begin{table*}[t]
\centering
\caption{Summary of spin-direction models used in this work and their distinct observational signatures on compact object binaries originating from isolated binary evolution.}
\label{tab:spin_summary}
\begin{tabular}{|p{6cm}|p{9cm}|}
\hline
Spin Model & Expected Observational Signatures \\ \hline
\textbf{\Snat} \newline (Spin inherited from progenitor rotation) & Spins are preferentially aligned with the orbital angular momentum, leading to positive $\chi_{\rm eff}$. Misalignment arises primarily due to natal kicks and increases with kick magnitude. \\ \hline
\textbf{\Siso} \newline (Isotropic spins) & Spin directions are randomly oriented relative to the orbit, resulting in a uniform distribution in $\cos\theta$ and a symmetric distribution of $\chi_{\rm eff}$ around zero. \\ \hline
\textbf{\Spar} \newline (Spins aligned with natal kicks) & Produces a distinctive double-peaked distribution in spin--orbit misalignment angles with a symmetric distribution of $\chi_{\rm eff}$ around zero. Primarily results in suppression of systems with  spins in the orbital plane. \\ \hline
\textbf{\Sperp} \newline (Spins perpendicular to natal kicks) & Also produces a double-peaked spin-tilt distribution, but with peaks at different angles than \Spar. Also produces a symmetric distribution of $\chi_{\rm eff}$ around zero. Tends to suppress spin aligned or anti-aligned with the orbital angular momentum. \\ \hline
\end{tabular}
\end{table*}
\end{nolinenumbers}

One of the least understood aspects of compact objects is the origin of their spins, especially those of black holes (BHs). The most popular theory suggests that compact objects inherit their spin from their parent star's angular momentum, specifically from the collapsing iron core of the progenitor star  \citep{Heger:2004qp, Ott:2006qp}. Since stars in binary systems tend to have their spins aligned with the orbital angular momentum, it is expected that compact objects should also have their spins preferentially aligned.
However, observations of compact object binaries in the field do not meet this expectation, showing no alignment with the orbital angular momentum. For instance, binary neutron star systems, where the young neutron star is visible as a radio pulsar, have been observed with large misalignment angles: $130.04 \pm 0.4^\circ$  for PSR J0737-3039B  and $104 \pm 9^\circ$ for  PSR J1906+0746 \citep{Breton:2008xy, Desvignes:2019uxs}. Furthermore, the X-ray binaries Cyg X-1~\citep{Zdziarski:2023ygh}, Cyg X-3~\citep{Zdziarski:2018dms}, and MAXI J1820+070~\citep{Poutanen:2021sag} also exhibit large spin misalignments of $20^\circ$--$30^\circ$, $37^\circ$, and $>40^\circ$, respectively. These observations show that compact objects might not inherit their spins from stars. Asteroseismic observations also support this conclusion. They show efficient angular momentum transport within massive stars, which slows core rotation prior to collapse \citep{Marques:2012eb, Cantiello:2014uja, Fuller:2019sxi, 2019A&A...626L...1E, Takahashi:2020jbi}. If the collapsing core has little angular momentum, then other mechanisms must be responsible for imparting spin to the resulting compact object.\footnote{We note that the spin magnitudes inferred for BHs in X-ray binaries are disputed and may be systematically overestimated \citep[see][for a recent review]{Zdziarski:2025ozs}; this concerns the inferred magnitudes rather than the geometric misalignment measurements quoted above.}

Given the growing evidence that compact object spins may not be simply inherited from their progenitor stars, it is important to consider alternative mechanisms that influence the direction of these spins. In this article, we focus on the implications of various nontraditional spin origins for the spin--orbit tilt of compact objects. These mechanisms will, in general, also affect the spin magnitudes, and we return to this briefly in Section~\ref{sec:pop}, but the orientation of the spin relative to the orbital angular momentum is our primary concern throughout. Specifically, we examine four distinct spin-direction models, each motivated by different theoretical and observational arguments:

\begin{enumerate}[itemsep=0pt, topsep=0pt]
    \item \Snat: Spins are inherited from the progenitor star (conventional model used in population synthesis and gravitational-wave astrophysics).
    \item \Siso: Spins are randomly oriented due to stochastic processes.
    \item \Spar: Spins are aligned with the direction of natal kicks.
    \item \Sperp: Spins are orthogonal to the direction of natal kicks.
\end{enumerate}
We elaborate on the motivation for each model below.

\subsection{Natal spins: \textnormal{\Snat}}
Most binary star systems display spins aligned with orbital angular momentum.\footnote{Note that binary stars with perfectly aligned spins are not always the norm, as demonstrated by the BANANA (Binaries Are Not Always Neatly Aligned) project. This project measured spin--orbit misalignments in close binary star systems, finding many notable exceptions to the spin alignment of binary stars~\citep{Albrecht:2007rf,Albrecht:2009bg,Albrecht:2010tr,Albrecht:2012fy,Albrecht:2014mxa,2022ApJ...933..227M,2024ApJ...975..149M}.} This alignment is often anticipated given that both stars originate from the same portion of a molecular cloud. If the angular momentum of a star is preserved during its collapse into a BH, the resulting BH should also have spin aligned with the orbital angular momentum.  As the prevailing model, it has been widely used in population synthesis, though it is often at odds with observations.

\subsection{Isotropic spins: \textnormal{\Siso}}
BHs may not necessarily inherit their spin from their parent star. Instead, they may acquire their spin through various mechanisms that could impart a spin in a random direction. Several such mechanisms have been proposed in the literature. For example, \citet{2014ApJ...796...17F, 2015ApJ...810..101F, Ma:2019cpr, McNeill:2020hbp} suggested that internal gravity waves excited during late-time burning phases or envelope convection could spin up nonrotating core-collapse (CC) progenitors. This stochastic spin-up leads to a Maxwellian distribution of precollapse core spin magnitudes and isotropic spin directions that are unrelated to the envelope's spin. Another possibility is that during a massive star's collapse without an energetic explosion, some of its hydrogen envelope may fall onto the newborn BH. The infalling convective shells in the star have random velocity fields that result in a finite angular momentum at each shell \citep{Gilkis:2014rda, 2016ApJ...827...40G, Quataert:2018gnt, Antoni:2021xzs, Antoni:2023yxs}. The accretion of this convective envelope can cause the BH to spin (with a final spin parameter of up to 0.8), despite the progenitor star not rotating.  \citet{Tauris:2022ggv} has also suggested that the BH spin axis may get tossed during the CC and might not be necessarily aligned with the orbital angular momentum  \citep[see also][for neutron star spin tossing]{Farr:2011gs}, potentially affecting the observed distribution of binary black hole (BBH) mergers by the LIGO-Virgo-KAGRA (LVK) Collaboration.  In CC simulations, transfer of angular momentum to a neutron star can also happen via stochastic accretion downflows  during the postbounce accretion phase, leading to the random orientation of neutron star spins~\citep{Wongwathanarat:2012zp}.

\subsection{BH spins parallel to natal kicks: \textnormal{\Spar}}
Spins of young pulsars are often observed to be aligned with their proper motion~\citep{Kramer:2003au, Johnston:2005ka, Wang:2005jg, Ng:2006vh, Johnston:2007gx, Chatterjee:2009ac, Noutsos:2012dt, Noutsos:2013ce, Mandel:2022sxv}. These proper motions are widely interpreted as the result of natal kicks, typically ranging from $O(10\ \kms)$ to $O(1000\ \kms)$ \citep{Hobbs:2005yx, Faucher-Giguere:2005dxp, Igoshev:2021bxr, Stevenson:2022hmi, ODoherty:2023vhq, Kapil:2022blf, Mandel:2022sxv}, imparted by asymmetric mass ejection during supernova events~\citep{1987IAUS..125..255W, Bisnovatyi-Kogan:1993ljj, 1994A&A...290..496J, Burrows:1995bb, Fryer:2005sz, Kusenko:2008gh, Sagert:2007as}. The observed alignment with spin axes suggests a possible physical connection between the kick mechanism and the spin direction. Since most observations can only measure the transverse velocity of pulsars, the alignment between the natal kicks and spin direction is inferred in a two-dimensional projection on the plane of the sky.
However, recent observations using the Five-hundred-meter Aperture Spherical Radio Telescope have shown that PSR J0538+2817 has almost perfect 3D alignment between its spin axis and velocity vector to within approximately $6^\circ$~\citep{2021NatAs...5..788Y}. \citet{Janka:2016nak} proposed three ways that can lead to this spin-kick alignment in neutron stars: spiral motions of the standing accretion shock instability (SASI), large-scale velocity and density perturbations in convective burning shells, and dipolar neutrino-emission asymmetry associated with the lepton-emission self-sustained asymmetry phenomenon. According to \citet{Janka:2021deg}, the natal kick causes the neutron star to be increasingly offset from the explosion center, leading to one-sided accretion of tangential vortex flows in the fallback matter, ultimately leading to spin-kick alignment \citep[see][for counterargument]{Muller:2023byy}. In simulations by \citet{Muller:2017hht}, the protoneutron star, which initially gained angular momentum in a direction perpendicular to the kick after shock revival, shows a trend toward spin-kick alignment as the angle between the spin axis and kick decreases from $80^\circ$ to $40^\circ$. Although longer simulations are required to confirm whether this trend continues toward even smaller misalignment, the observed trend suggests the possible existence of a systematic alignment mechanism. CC simulations by \citet{Burrows:2023ffl}  also discovered a potential correlation or anti-correlation between the spin and natal kicks of compact objects.  They observed that the explosion and accretion tend to occur roughly perpendicular to each other, with accretion spinning up the periphery of the protoneutron star. Consequently, the net spin vector aligns either parallel or antiparallel to the kick direction. The mechanisms discussed above also motivate a different choice of convention. In population synthesis and gravitational-wave studies, spin-kick alignment is almost always implemented by treating the spin axis as given --- inherited from the progenitor and aligned with the orbital angular momentum --- and drawing the natal kick along it \citep{Kalogera:1999tq, Willems:2004vf, Willems:2004kc, Gerosa:2013laa, Callister:2020vyz, Gompertz:2021xub}. The physical picture above reverses this: the asymmetry that generates the kick also generates the spin, and none of these mechanisms requires the compact object to possess any angular momentum beforehand --- several are in fact motivated by the expectation that the precollapse core rotates slowly. We therefore adopt the opposite convention throughout this work: the natal kick sets the preferred direction, and the BH spin is oriented with respect to it.

\subsection{BH spins perpendicular to natal kicks: \textnormal{\Sperp}}
Although pulsar observations have shown spin-kick alignment, and various studies propose hydrodynamical mechanisms to explain it, most 3D CC simulations show no indication of spin-kick alignment \citep{Wongwathanarat:2012zp, Gessner:2018ekd, Chan:2020lnd, Stockinger:2020hse, Powell:2020cpg}, with the exception of \citet{Burrows:2023ffl} finding indications of alignment. Instead, \citet{Chan:2020lnd} showed that an asymmetric explosion with anisotropic and partial fallback can result in a BH spin of approximately $0.25$, with the kick and spin vectors perpendicular to each other  \citep[see Figure~10 in][]{Chan:2020lnd}. Similar trends were identified by \citet{Stockinger:2020hse} with perpendicular kick and spin, while \citet{Powell:2020cpg} found a large angle of $70^\circ$ between the two vectors. \citet{Spruit:1998sg} have also suggested an off-center kick scenario where the remnant receives a momentum impulse $\mathbf{v_k}$ that is misaligned with the center of mass by $\mathbf{r_{\rm offset}}$, resulting in the spin vector being perpendicular to the kick ($\mathbf{r_{\rm offset}}\times\mathbf{v_k}$). \citet{Farr:2011gs} used off-center kicks to explain the large misalignments between spin and the orbital angular momentum of the binary pulsar PSR J0737-3039. Such off-center explosions can be explained when the explosion is launched out of a SASI-active phase  \citep{Blondin:2006yw, Fernandez:2010db, Guilet:2013bxa, Kazeroni:2015qca, Kazeroni:2017fup}, as observed by \citet{Janka:2016nak}. \citet{Muller:2023byy} showed that one-sided fallback accretion onto a compact object could also significantly contribute angular momentum perpendicular to the kick velocity.

In gravitational-wave astrophysics and binary population synthesis studies, it is commonly assumed that BHs formed through isolated binary evolution will have spins aligned with the binary's orbital angular momentum, as predicted by the \Snat model.
As a result, it has been widely believed that BBH mergers detected by LVK could be distinguished by their spin orientations~\citep{Gerosa:2013laa, Rodriguez:2016vmx, Vitale:2015tea, Talbot:2017yur, Farr:2017uvj, Stevenson:2017dlk, Farr:2017gtv, Gerosa:2018wbw, Tiwari:2018qch, Fernandez:2019kyb, Baibhav:2022qxm}. Specifically, dynamically assembled BBHs are expected to have spins that are isotropically distributed. In contrast, BBHs formed through isolated binary evolution are assumed to have spins aligned with the orbital angular momentum, reflecting the presumed inheritance of spin from the progenitor stars.
However, this view is overly simplistic. In this article, we explore how different mechanisms, other than simple spin inheritance from progenitors, affect the spin orientations of compact objects in binaries. We find that the conventional belief that isolated binary evolution can only produce BBHs with preferentially aligned spins is not necessarily true. Traditionally, gravitational-wave events with large spin--orbit misalignments, such as GW190412, GW190521, and GW191109, have been interpreted as originating from dynamically assembled binaries in dense stellar environments~\citep{Gerosa:2020bjb, Rodriguez:2020viw, Tagawa:2020dxe, Baibhav:2020xdf, Liu:2020gif, Gerosa:2021hsc, Mapelli:2021syv, Gerosa:2021mno, Baibhav:2021qzw, Mapelli:2021gyv, Fishbach:2022lzq, Li:2022mck, Zhang:2023fpp}. However, our work challenges this notion. We propose that such binaries could also form in the field, where compact object spins are not necessarily aligned with the orbital angular momentum. Furthermore, our work suggests that the identification of precession or an effective inspiral spin parameter \chieff less than zero, which has traditionally been considered unlikely in isolated binary mergers, is a distinct possibility. This challenges the ability to distinguish between field binaries and dynamically assembled binaries based solely on spin orientations.

In Section~\ref{sec:LS}, we present the mathematical framework used to calculate spin--orbit misalignments under each of the four spin-direction models. We explore how CC and natal kicks influence spin directions in these scenarios. Among these, the \Snat model is well studied in the literature, while the \Siso model, which assumes random spin directions, is conceptually straightforward. The framework for \Spar and \Sperp models, which introduce systematic alignments of spin with or perpendicular to the kick direction, is novel and forms the core of our new contributions. The main characteristics of each spin-direction model are summarized in Table~\ref{tab:spin_summary}.
In Section~\ref{sec:pop}, we apply these models to a simplified binary population to predict the distribution of spin--orbit tilt angles. Our aim is not to perform a detailed or comprehensive population synthesis, but rather to explore the broad effects of key parameters, such as natal kick magnitudes, evolutionary pathways, and tidal interactions, on the resulting spin misalignment distributions.  We analyze the impact of each spin-direction model on gravitational-wave (GW) observables: primarily the spin--orbit tilts and the effective inspiral spin, and secondarily the spin magnitudes. In Section~\ref{sec:LVK}, we compare these predictions with LVK observations at the individual-event level. Finally, we provide our conclusions in Section~\ref{sec:conclusions}.

\section{Spin-orbit misalignments}
\label{sec:LS}
\begin{figure}[t]
   \includegraphics[width=\columnwidth]{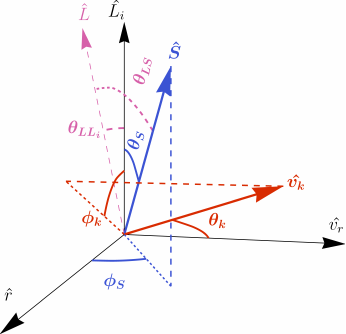}
    \caption{Angles involved in studying natal kicks. The binary system components are initially separated along the unit vector $\mathbf{\hat{r}}$, with a relative velocity of $\mathbf{v_0}$, before the CC. The BH progenitor has a spin given by $(\thetaS, \phiS)$ (highlighted in red) and receives a natal kick along the unit vector $\mathbf{\hat{v_k}}$ given by angles $(\thetaK, \phiK)$ (highlighted in blue). Due to the natal kick, the orbital angular momentum tilts by an angle $\thetaLL$ from the $\mathbf{L_i}$ to  $\mathbf{L}$ (highlighted in pink). The change in the orbit tilt causes the spin--orbit misalignment to change from $\thetaS$ to $\thetaLS$.
    }
    \label{fig:angles}
\end{figure}
We set up a reference frame to analyze the spin--orbit dynamics as illustrated in Figure~\ref{fig:angles}.
We assume a binary star system on a circular orbit with mass $M_i$, initial separation $a_i \mathbf{\hat{r}}$ and orbital angular momentum along $\mathbf{L_i}$.  A CC causes one of the stars to receive a natal kick $\mathbf{v_k}$, where the kick direction is defined using two angles, namely the polar angle $\thetaK$ and the azimuthal angle $\phiK$. The polar angle $\thetaK$ represents the angle between the kick velocity vector $\mathbf{v_k}$ and the pre-CC orbital velocity vector $\mathbf{v_0}$, and the azimuthal angle $\phiK$ is the angle subtended in the $\mathbf{L_i}-\mathbf{r}$ plane, where $\phiK=0$ implies kicks parallel to the initial angular momentum $\mathbf{L_i}$.

The CC decreases the total mass of the binary system from $M_i$ to $M_f$ and alters the velocity of the exploding star from $\mathbf{v_0}$ to $\mathbf{v_0} + \mathbf{v_k}$. Applying energy and angular momentum conservation to the binary system before and after the CC, we can determine the final values of semimajor axis ($a_f$) and eccentricity ($e$) as follows~\citep{1995MNRAS.274..461B, Kalogera:1996rm, Kalogera:1999tq}:

\begin{align}\label{eq:ae}
\frac{a_f}{a_i}&= \frac{\beta}{ \left[ 2(\beta-1) +1-u_k^2 - 2 u_k \cos\thetaK \right]},\\
1 - e^2 &= \frac{2 \beta -u_k^2 \sin ^2\thetaK-(u_k \cos \thetaK +1)^2 }{\beta ^2}\nn\\
&\times \left(u_k^2 \sin ^2\thetaK  \cos ^2\phiK+(1+u_k \cos \thetaK)^2\right)
\end{align}

where $\beta$ is the ratio of the final mass $M_f$ to the initial mass $M_i$, and $u_k$ is the kick magnitude normalized to the circular orbital velocity before the explosion:

\begin{align}
u_k=\frac{v_k}{v_0}=v_k\sqrt{\frac{a_i}{G M_i}}.
\end{align}

When $e>1$, the CC will disrupt the binary system. This occurs when the kick velocity $\mathbf{v_k}$ has a large component along the initial orbital velocity vector $\mathbf{v_0}$. The maximum value of $\cos \thetaK$, beyond which the binaries get disrupted, can be expressed as~\citep{1995MNRAS.274..461B, Kalogera:1996rm, Kalogera:1999tq}

\begin{align}
\cos \thetaK < \cos \thetaK^{\rm max} \equiv \frac{2 \beta -u_k^2-1}{2 u_k}
\end{align}

The probability of a binary not getting disrupted is $(1+\cos \thetaK^{\rm max})/2$. It should be noted that binaries do not disrupt when $u_k<\sqrt{2\beta}-1$, while all binaries will disrupt if $u_k>\sqrt{2\beta}+1$~\citep{1995MNRAS.274..461B}.

The  binary separation due to a CC changes between~\citep{1995MNRAS.274..461B, Kalogera:1996rm, Kalogera:1999tq}

\begin{equation}
\frac{a_f}{a_i}\in \Bigl[ \frac{\beta }{2 \beta -(1-u_k)^2} ,\infty\Bigr)
\end{equation}

The CC also causes the orbital plane of the binary to tilt by an angle $\thetaLL$~\citep{Kalogera:1999tq},
\begin{eqnarray}
\cos \thetaLL &=& \frac{u_k \cos \thetaK +1}{\sqrt{u_k^2 \sin ^2\thetaK \cos ^2\phiK +(u_k \cos \thetaK+1)^2}}\label{eq:costhetaLL}\\
\sin \thetaLL &=& -\frac{u_k \sin\thetaK \cos\phiK}{\sqrt{u_k^2 \sin^2\thetaK \cos^2\phiK +(u_k \cos\thetaK+1)^2}}\nn
\end{eqnarray}

In the case of low-magnitude kicks, the orbital tilts are limited to small values~\citep{Kalogera:1999tq}.
As the natal kick magnitude increases, the range of allowed tilt angles in BBH systems expands, and the peak of the angle distribution shifts toward higher values. For $1\leq u_k<\sqrt{2\beta+1}$, the full range of orbital tilt angles, from alignment ($\cos\thetaLL=1$) to antialignment ($\cos\thetaLL=-1$), becomes possible. When $u_k>\sqrt{2\beta+1}$, the surviving binaries have tilt angles greater than $90^\circ$, leading to retrograde post-CC orbits. The allowed range of orbital tilts for a given $u_k$ is
\begin{align}\label{eq:cosLLRange}
\cos\thetaLL \in &\nn\\&\hspace{-1cm} \begin{dcases}
\bigl(\sqrt{1 - u_k^2},1\bigr] & u_k<1\\
\bigl[-1,1\bigr] & 1\le u_k<\sqrt{2 \beta +1}\\
\biggl[-1,\frac{1-u_k^2+2 \beta }{ \sqrt{8\beta }}\biggr] & \sqrt{2 \beta +1}\le u_k<\sqrt{2 \beta }+1\\
\text{disrupted} & u_k\ge\sqrt{2 \beta }+1\,.
\end{dcases}
\end{align}

The direction of the spin vector $\mathbf{S}$ of the collapsing star can be specified by two angles, namely the angle $\thetaS$ between the spin vector $\mathbf{S}$ and the orbital angular momentum vector $\mathbf{L_i}$, and the angle $\phiS$ between the projection of the spin vector $\mathbf{S}$ onto the orbital plane and the separation vector $\hat{\mathbf{r}}$ between the binary members. The CC causes the angle between the binary's spin vector $\mathbf{S}$ and its orbital angular momentum vector $\mathbf{L}$ to change from $\thetaS$ to $\thetaLS$, where

\begin{align}\label{eq:costhetaLS}
\cos \thetaLS = \cos\thetaS\cos\thetaLL + \sin\thetaS\sin\thetaLL\sin\phiS
\end{align}

Our goal is to understand how different physical mechanisms that impart spins to BHs also impact post-CC spin--orbit misalignment. Each of these mechanisms generates distinct descriptions of the BH spin direction, represented by $(\thetaS, \phiS)$, which will have varying effects on the distributions of $\thetaLS$. Next, we explore four potential scenarios of spin angles and analyze how they shape the distributions of $\thetaLS$ and study how these mechanisms can be observed through the unique signatures they leave on the population of BH mergers.

\subsection{Misalignment distribution for \textnormal{\Snat}}
 The distribution of spin--orbit misalignment for natal spins initially aligned with orbital angular momentum  has been extensively explored in~\citet{Kalogera:1999tq, Rodriguez:2016vmx, OShaughnessy:2017eks, Rodriguez:2018jqu, Wysocki:2017isg, Atri:2019fbx, Salvesen:2020nkm, Zhu:2021zxj, Callister:2020vyz, Steinle:2020xej, Steinle:2022rhj, Gompertz:2021xub}.
If the spin vector $\mathbf{S}$ is aligned with $\mathbf{L}$ before the CC ($\thetaS = 0$), the tilt of the orbital plane equals the misalignment of the exploding star's spin ($\thetaLS = \thetaLL$).

\begin{figure*}
   \includegraphics[width=\linewidth]{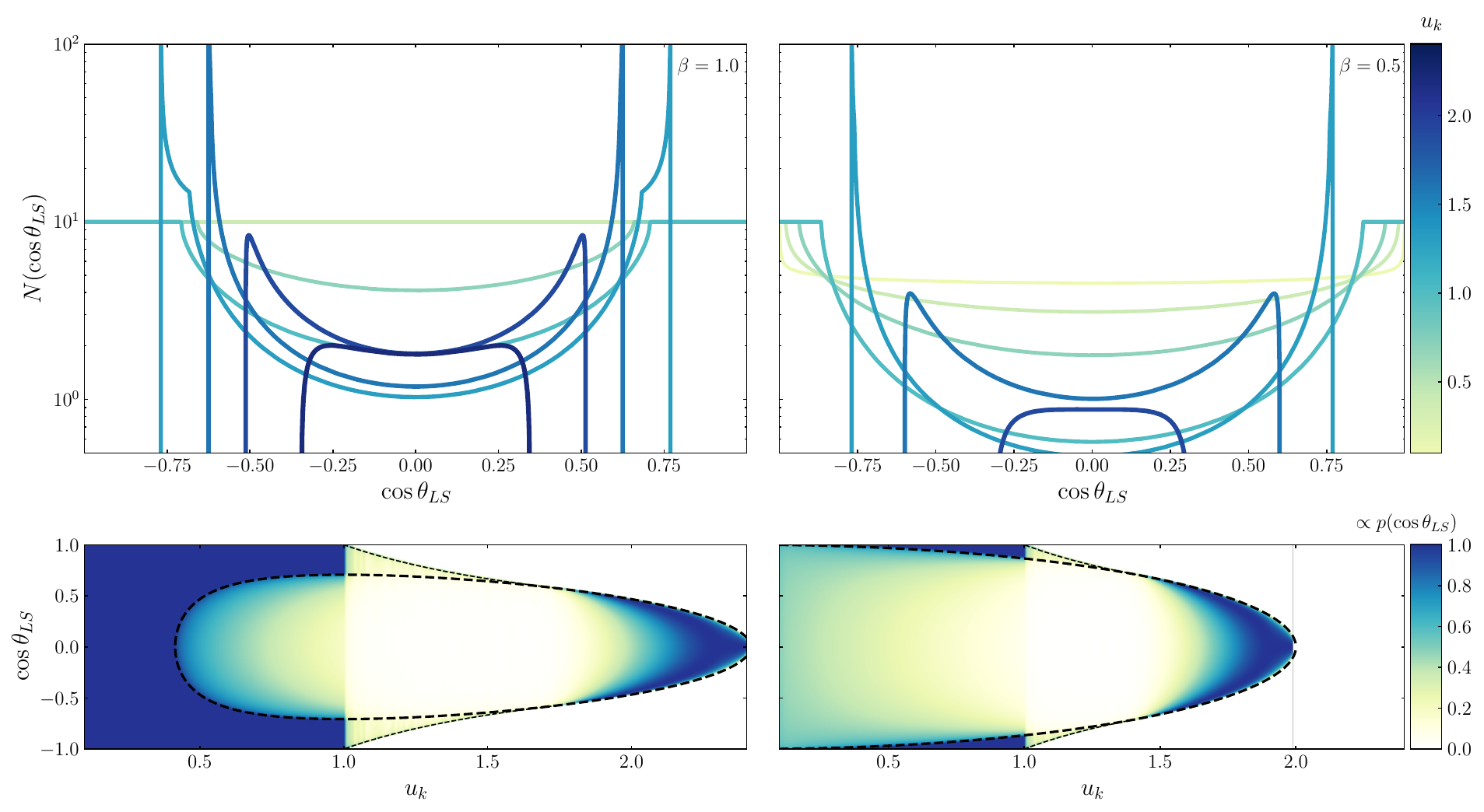}
    \caption{Distributions of the spin--orbit misalignment $\cos\thetaLS$ for the \Spar model (in which the BH acquires spin aligned with the natal kick direction). Top panels: histograms $N(\cos\thetaLS)$ of  $\cos\thetaLS$ for different values of $u_k$ (natal kick magnitude normalized by orbital velocity), ranging from $0.1$ (light green) to $2.2$ (dark blue) in increments of $0.3$. Left and right panels correspond to $\beta=1$ (no mass loss) and $\beta=0.5$ ($50\%$ binary mass loss), respectively. Lower panels: probability density functions of $\cos\thetaLS$, normalized by the maximum value, as a function of $u_k$ and $\cos\thetaLS$. The dashed lines correspond to Equation~(\ref{eq:ParReducedLS}), highlighting the regions where spin--orbit misalignments are suppressed for small kicks and favored for large kicks. Dotted lines show the location of the peaks at $\cos\theta_{\rm LS} = \pm 1/u_k$. }
    \label{fig:CosThetaLS_Par}
\end{figure*}

\subsection{Misalignment distribution for \textnormal{\Siso}}
For BHs that acquire their spin through stochastic processes, there is no reason to expect a preferred spin orientation relative to the orbital angular momentum. As a result, we assume an isotropic spin--orbit misalignment, for which $\cos\theta_{\rm LS}$ is uniformly distributed over $[-1,1]$.

\begin{figure*}[t]
   \includegraphics[width=\linewidth]{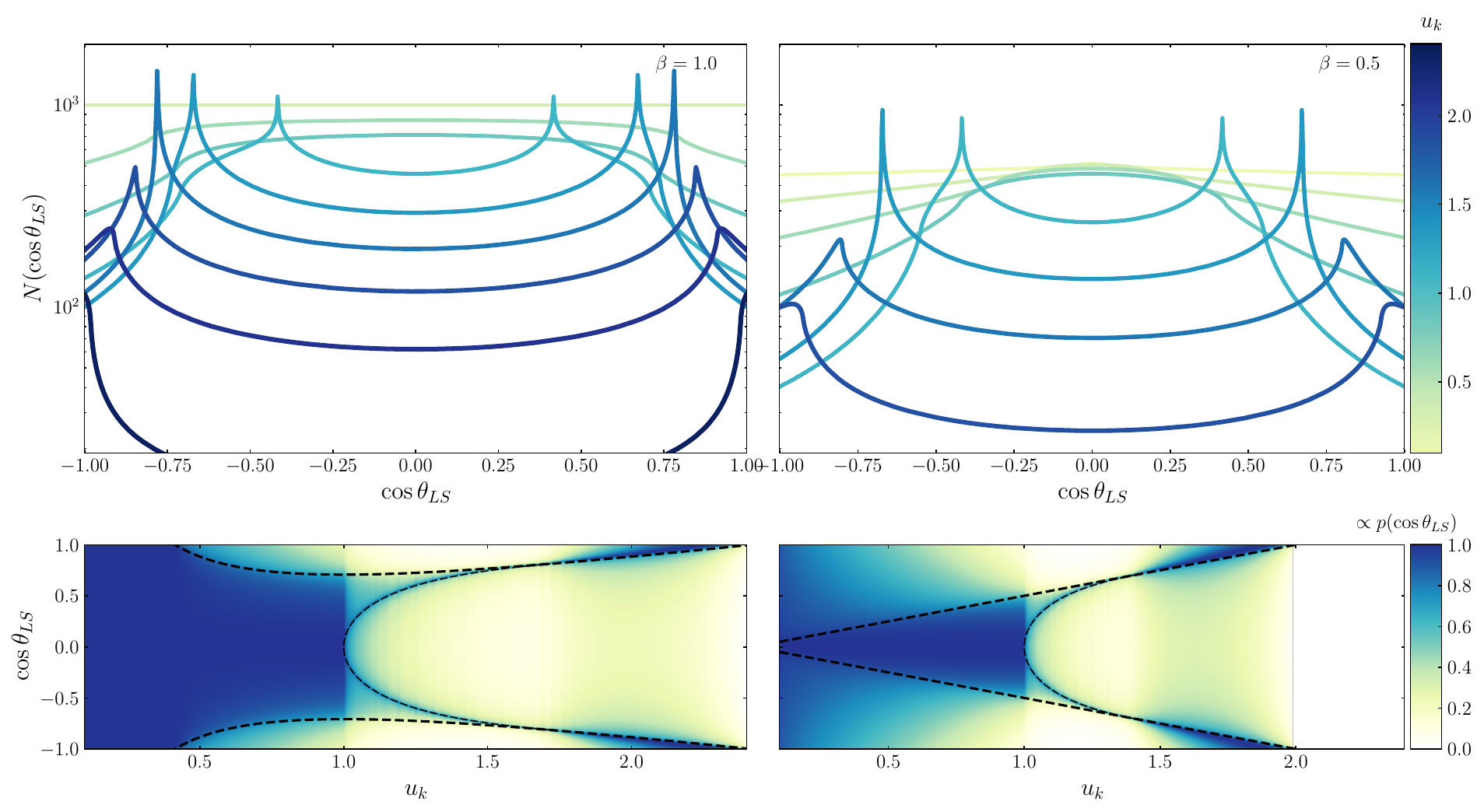}
    \caption{Distributions of the spin--orbit misalignment $\cos\thetaLS$ for the \Sperp model (in which the BH acquires spin perpendicular to the natal kick direction). Top panels: histograms $N(\cos\thetaLS)$ of  $\cos\thetaLS$ for different values of $u_k$ (natal kick magnitude normalized by orbital velocity), ranging from $0.1$ (light green) to $2.2$ (dark blue) in increments of $0.3$. Left and right panels correspond to $\beta=1$ (no mass loss) and $\beta=0.5$ ($50\%$ binary mass loss), respectively. Lower panels: probability density functions of $\cos\thetaLS$, normalized by the maximum value, as a function of $u_k$ and $\cos\thetaLS$. The dashed lines correspond to Equation~(\ref{eq:PerpReducedLS}), highlighting the regions where spin--orbit misalignments are suppressed for small kicks and favored for large kicks. Dotted lines show the location of the peaks at $\cos\theta_{\rm LS} = \pm\sqrt{u_k^2-1}/u_k$.
    }
    \label{fig:CosThetaLS_Perp}
\end{figure*}

\subsection{Misalignment distribution for \textnormal{\Spar}}

In the \Spar model, we assume that the natal kick direction is isotropic and that the compact object is spun up parallel to the kick through one of the mechanisms discussed in Section~\ref{sec:intro}.

In this case $\mathbf{S}\parallel \mathbf{v_k}$, and the angle subtended by the spin with the initial orbital angular momentum is
\begin{align}
\cos\thetaS= & \sin\thetaK \cos\phiK \nn\\
\sin\phiS = & \frac{\cos\thetaK}{\sqrt{1-\sin^2\thetaK\cos^2\phiK}}
\end{align}
or using Equation~(\ref{eq:costhetaLS}), the tilt angle between final orbital angular momentum and spin would be
\be
\cos\thetaLS=\frac{\sin \thetaK \cos \phiK}{\sqrt{u_k^2 \sin ^2\thetaK \cos ^2\phiK+(u_k \cos \thetaK+1)^2}}
\ee

To study the distribution of spin--orbit misalignment, we perform Monte Carlo experiments involving $10^6$ binaries, where we vary the value of $u_k$ from 0 to $\sqrt{2\beta}+1$. We assume two extreme mass loss scenarios, namely, $\beta=0.5$ and $\beta=1$. Additionally, we assume that the natal kicks have isotropic directions, which we sample uniformly in $\cos\thetaK$ and $\phiK$. The results of the Monte Carlo experiments are shown in Figure~\ref{fig:CosThetaLS_Par}. The top panels of the figure show the histograms for $\cos\thetaLS$ for $u_k$ ranging from $0.1$ (light green) to $2.2$ (dark blue) in increments of $0.3$ for $\beta=1$ (left panel) and $\beta=0.5$ (right panels). In the lower panel, we plot the probability density functions of $\cos\thetaLS$ as a function of $u_k$. We normalize these probability density functions such that the maximum value is 1. We expect that since the kick directions are isotropic and the spins are parallel to the kicks, the spins would also be isotropic. However, not all kick directions are allowed as they disrupt the binary. This effect is reflected in the distributions in Figure~\ref{fig:CosThetaLS_Par}.
For $u_k<\sqrt{2\beta}-1$, no binary is disrupted, and therefore, the distribution of spin--orbit tilts is isotropic. However, at $u_k>\sqrt{2\beta}-1$, the binaries begin to get disrupted, particularly if they have a large kick contribution along the orbital velocities ($\cos\thetaK>\cos\thetaK^{\rm max}$). This effect impedes binaries from having large in-plane spins. For $\sqrt{2\beta}-1<u_k<1$, there is a reduced number of systems with tilts in the range

\be\label{eq:ParReducedLS}
\left|\cos\thetaLS\right|<\sqrt{\frac{1}{8\beta }\left(4 \beta-u_k^2+2 -\frac{(1-2 \beta )^2}{u_k^2}\right)}
\ee

In Figure~\ref{fig:CosThetaLS_Par}, for $u_k>1$, we observe interesting features. Specifically, $\cos\thetaLS$ is bounded between

\be\label{eq:ParPeak}
-\frac{1}{u_k} \le \cos\thetaLS \le \frac{1}{u_k}\,,
\ee

 with the upper and lower bounds occurring at $\cos\thetaK=-1/u_k$ and $\cos\thetaLL=0$. We see very sharp peaks in the distribution at these extrema.
 Moreover, for $u_k>\sqrt{2\beta+1}$, we observe an enhanced number of systems with spin--orbit tilts in the region given by Equation~(\ref{eq:ParReducedLS}). This reversal in trend can be understood by considering the absence of systems with spins in the pre-CC orbital plane. For $u_k<1$, the angular momentum tilts are small, so the change in pre-CC and post-CC planes is also small. Therefore, there is a lack of systems with spins in post-CC planes. On the other hand, for $u_k>\sqrt{2\beta+1}$, the orbit tilts by at least $90^\circ$ (see Equation~(\ref{eq:cosLLRange})). Consequently, the spins that were previously in-plane and disfavored are now parallel to the orbital angular momentum. Similarly, spins that were initially aligned with the orbital angular momentum and favored are now in the post-CC orbital plane.

An alternative way to understand some of these features is by examining the relationship between $\cos\thetaLS$ and $\cos\thetaLL$:
\begin{equation}\label{eq:ThetaLSLL_Par}
\left(u_k \cos\thetaLS\right)^2 + \left(\cos\thetaLL\right)^2=1.
\end{equation}
 The allowed range of $\cos\thetaLL$ depends on the values of $u_k$ and $\beta$ according to Equation~(\ref{eq:cosLLRange}), thereby also restricting the possible values of $\cos\thetaLS$. When $u_k<1$, $\cos\thetaLL$ is positive, i.e., the orbit cannot tilt by more than $90^\circ$; therefore, $\cos\thetaLS$ can assume any value in $[-1,1]$. As $u_k$ increases beyond $1$, the maximum value of $|\cos\thetaLS|$ decreases to $1/u_k$. However, for $u_k>\sqrt{2\beta+1}$, the tilt of the orbit is more than $90^\circ$, and the minimum value of $\cos\thetaLL$ is given by Equation~(\ref{eq:cosLLRange}). This further restricts the allowed values of $\cos\thetaLS$ according to Equation~(\ref{eq:ThetaLSLL_Par}).

 To summarize, in the model \Spar where BHs receive their spins parallel to the kicks, there are several important features of spin--orbit misalignment, including:
\begin{itemize}
    \item For $u_k<\sqrt{2\beta}-1$, the spin--orbit misalignments are isotropic.
    \item For $\sqrt{2\beta}-1<u_k< \sqrt{2\beta+1}$, there is a reduced number of systems with the spin--orbit misalignment given by Equation~(\ref{eq:ParReducedLS}).
    \item For $u_k>1$, sharp peaks appear at the extrema of $\cos\thetaLS=\pm 1/u_k$.
    \item For $u_k>\sqrt{2\beta+1}$, all the tilts are in the region given by Equation~(\ref{eq:ParReducedLS}).
\end{itemize}

\subsection{Misalignment distribution for \textnormal{\Sperp}}
Given a natal kick $\mathbf{\hat{v}_k}$, let us define the spin plane by two unit vectors  as $\mathbf{e_{1}}= \sin\phiK \mathbf{\hat{L}}- \cos\phiK \mathbf{\hat{r}}$ (such that $\mathbf{e_{1}}\cdot\mathbf{\hat{v}_k}=0$ ) and $\mathbf{e_{2}}=\mathbf{\hat{v}_k}\cross \mathbf{e_{1}}$. We can then define the spin direction as
\be
\mathbf{\hat{S}}= \sin\phi_r\ \mathbf{e_{1}} + \cos\phi_r\ \mathbf{e_{2}}
\ee
Thus, $\mathbf{\hat{S}}$ is given by three angles: $\phiK$ and $\thetaK$ that decide the normal vector $\mathbf{\hat{v}_k}$ to the spin plane, while $\phi_r$ decides the orientation of the spin in this plane. The angle subtended by the spin with the initial orbital angular momentum is given by
\begin{align}
\cos\thetaS= & \cos\thetaK\cos\phi_r\cos\phiK +\sin\phi_r\sin\phiK  \nn\\
\sin\phiS= & \nn\\
& \hspace{-1.3cm}\frac{-\sin \thetaK \cos \phi_r}{\sqrt{(\cos \thetaK \cos \phi_r \sin \phiK-\sin \phi_r \cos \phiK)^2+(\sin\thetaK \cos\phi_r)^2}}
\end{align}
or using Equation~(\ref{eq:costhetaLS}), the tilt angle between final orbital angular momentum and spin would be
\begin{align}
\cos\thetaLS= & \nn\\
& \hspace{-1.4cm}\frac{(u_k+\cos\thetaK)\cos\phi_r\cos\phiK+(1+u_k\cos\thetaK)\sin\phi_r\sin\phiK}{\sqrt{u_k^2\sin^2\thetaK\cos^2\phiK+(1+u_k\cos\thetaK)^2}}
\end{align}

To explore the distribution of spin--orbit misalignment, we perform Monte Carlo experiments by randomly sampling $10^6$ binaries with uniform values of $\cos\thetaK$, $\phiK$, and $\phi_r$. We vary the value of $u_k$ from 0 to $\sqrt{2\beta}+1$ while considering two extreme mass loss scenarios: $\beta=0.5$ and $\beta=1$. The results of this Monte Carlo are shown in Figure~\ref{fig:CosThetaLS_Perp}. The top panels of the figure show histograms of $\cos\thetaLS$ for $u_k$ ranging from $0.1$ (light green) to $2.2$ (dark blue) in increments of $0.3$ for $\beta=1$ (left panel) and $\beta=0.5$ (right panel). In the lower panel, we plot the probability density functions of $\cos\thetaLS$, normalized such that the maximum value is 1, as a function of $u_k$.

When $u_k<\sqrt{2\beta}-1$, no binary is disrupted, and therefore, the distribution of spin--orbit tilts is isotropic. However, at $u_k>\sqrt{2\beta}-1$, binaries begin to get disrupted, especially if they have a large kick contribution along the orbital velocities ($\cos\thetaK>\cos\thetaK^{\rm max}$). This effect impedes binaries from having spins aligned with the orbital angular momentum. For $\sqrt{2\beta}-1<u_k<1$, there is a reduced number of systems with tilts in the range
\be\label{eq:PerpReducedLS}
\left|\cos\thetaLS\right|>\frac{2 \beta +u_k^2-1}{u_k \sqrt{8 \beta} }
\ee

In Figure~\ref{fig:CosThetaLS_Perp}, we observe interesting features for $u_k>1$. Specifically, the distribution of $\cos\thetaLS$ has sharp peaks at
\be\label{eq:PerpPeak}
\cos\thetaLS =\pm \frac{\sqrt{u_k^2-1}}{u_k}\,,
\ee
corresponding to $\cos\thetaK=-1/u_k$ and $\phi_r=\phiK=0$.
Additionally, for $u_k>\sqrt{2\beta+1}$, there is an increase in the number of systems with tilts given by Equation~(\ref{eq:PerpReducedLS}). This reversal in trend compared to behavior at $u_k<1$ can be understood by considering the absence of systems with spins aligned with the initial orbital angular momentum. For $u_k<1$, the angular momentum tilts are small, resulting in a lack of systems with aligned spins. However, for $u_k>\sqrt{2\beta+1}$, the orbit tilts by at least $90^\circ$ (as shown in Equation~(\ref{eq:cosLLRange})). Consequently, the spins that were previously aligned with $\mathbf{L_i}$ and disfavored are now perpendicular to $\mathbf{L}$.
 To summarize, in the model \Sperp where BHs receive their spins perpendicular to the kicks, there are several important features of spin--orbit misalignment, including:
\begin{itemize}
    \item For $u_k<\sqrt{2\beta}-1$, the spin--orbit misalignments are isotropic.
    \item For $\sqrt{2\beta}-1<u_k< \sqrt{2\beta+1}$, there is a reduced number of systems with the spin--orbit misalignment given by Equation~(\ref{eq:PerpReducedLS}).
    \item For $u_k>1$, sharp peaks appear at the extrema of $\cos\thetaLS=\pm \sqrt{u_k^2-1}/u_k$.
    \item For $u_k>\sqrt{2\beta+1}$, tilt distributions are concentrated in the region given by Equation~(\ref{eq:PerpReducedLS}).
\end{itemize}

\begin{figure*}[t]
   \includegraphics[width=\textwidth]{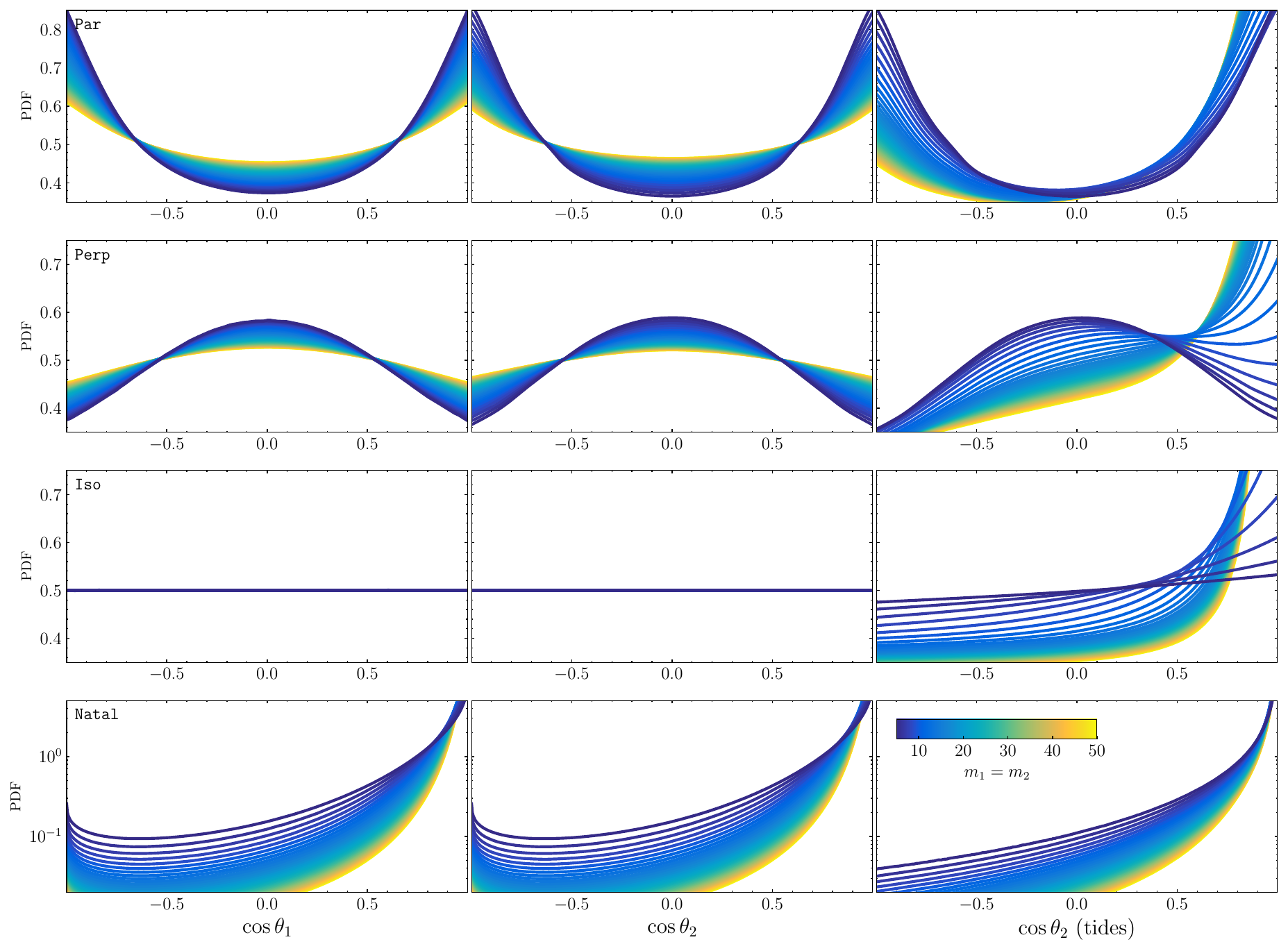}
    \caption{Probability density functions (PDFs) of spin--orbit tilts based on the noninteracting model for pre-CC2 orbital separations and high kicks for all BH masses: $p(\cos\theta_1)$ (left), $p(\cos\theta_2)$ assuming inefficient tides (middle), and $p(\cos\theta_2)$ with tidal spin-up (right) for the \Spar, \Sperp, \Siso, and \Snat models (from top to bottom) for BH masses ranging from $m_1=m_2=5 \Msun$ (in blue) to $50 \Msun$ (in yellow).}
    \label{fig:costheta_MVK_250_noninteracting}
\end{figure*}

\begin{figure}[t]
   \includegraphics[width=\columnwidth]{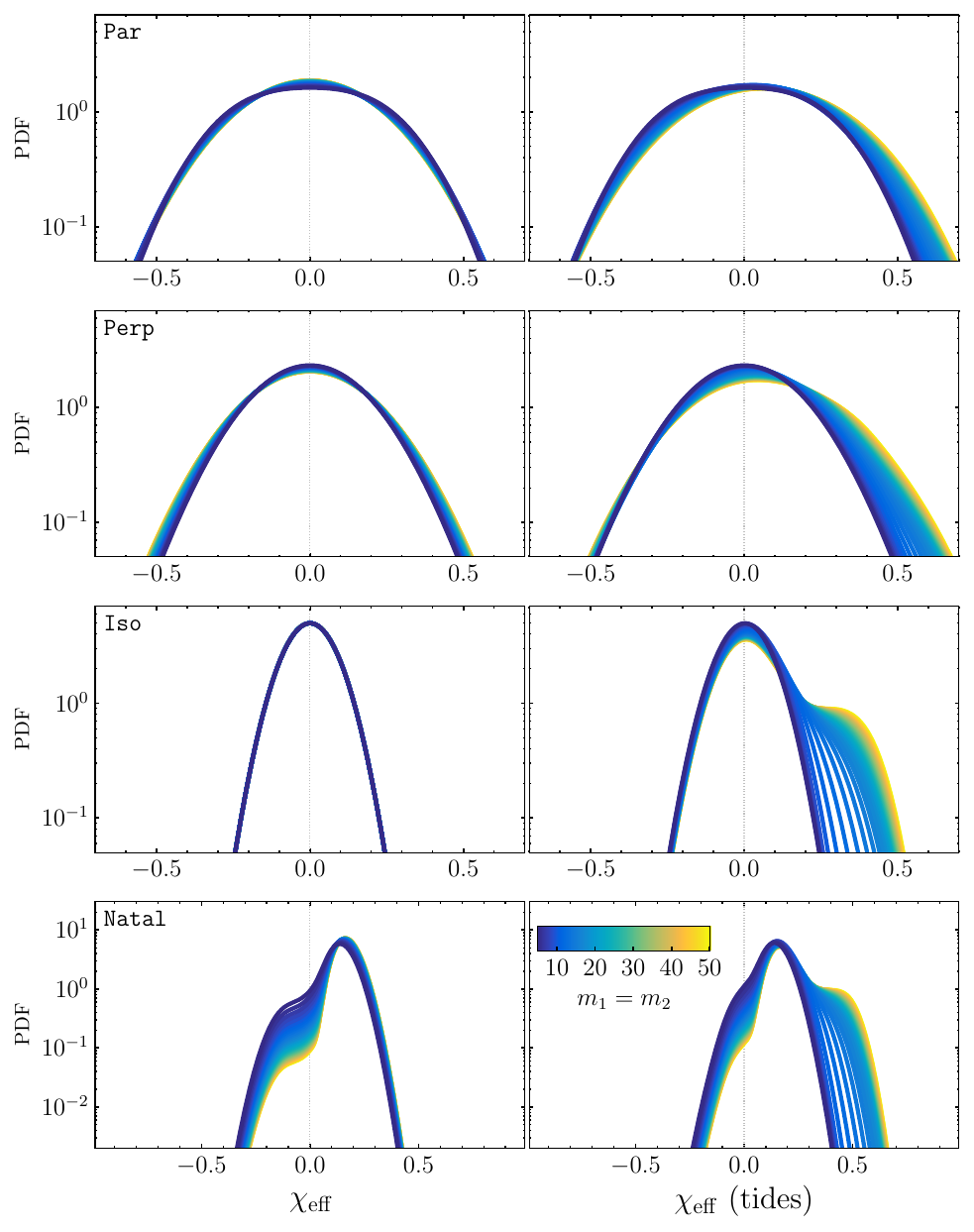}
    \caption{PDFs of \chieff based on the noninteracting model for pre-CC2 orbital separations and high kicks for all BH masses: $p(\chieff)$ assuming inefficient tides (left), and $p(\chieff)$ with tidal spin-up (right) for the \Spar, \Sperp, \Siso, and \Snat models (from top to bottom) for BH masses ranging from $m_1=m_2=5 \Msun$ (in blue) to $50 \Msun$ (in yellow).}
    \label{fig:chieff_MVK_250_noninteracting}
\end{figure}

\begin{figure*}[t]
   \includegraphics[width=\textwidth]{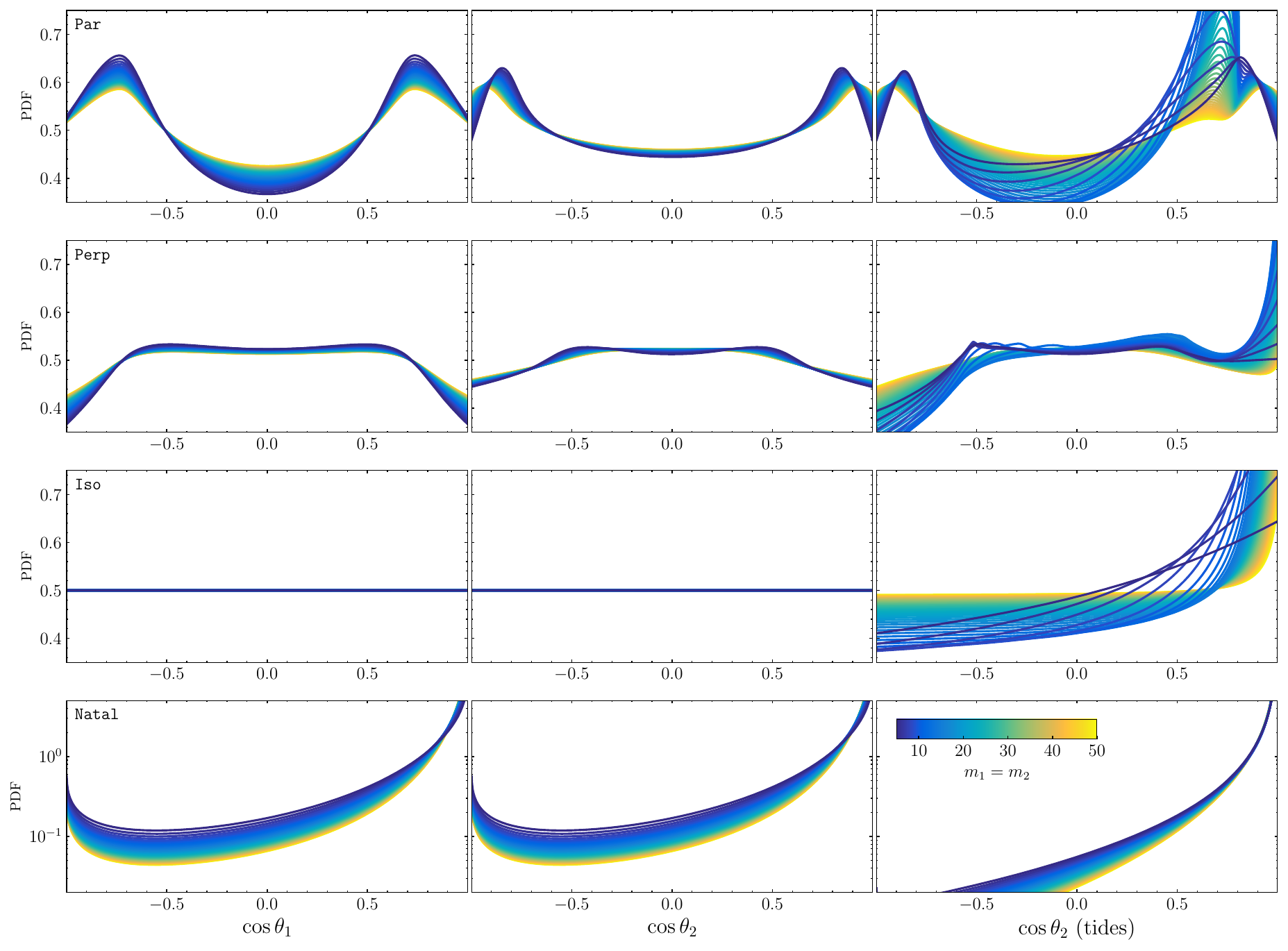}
    \caption{PDFs of spin--orbit tilts based on the interacting model for pre-CC2 orbital separations and high kicks for all BH masses: $p(\cos\theta_1)$ (left), $p(\cos\theta_2)$ assuming inefficient tides (middle), and $p(\cos\theta_2)$ with tidal spin-up (right) for the \Spar, \Sperp, \Siso, and \Snat models (from top to bottom) for BH masses ranging from $m_1=m_2=5 \Msun$ (in blue) to $50 \Msun$ (in yellow).}
    \label{fig:costheta_MVK_250_interacting}
\end{figure*}

\begin{figure}[t]
   \includegraphics[width=\columnwidth]{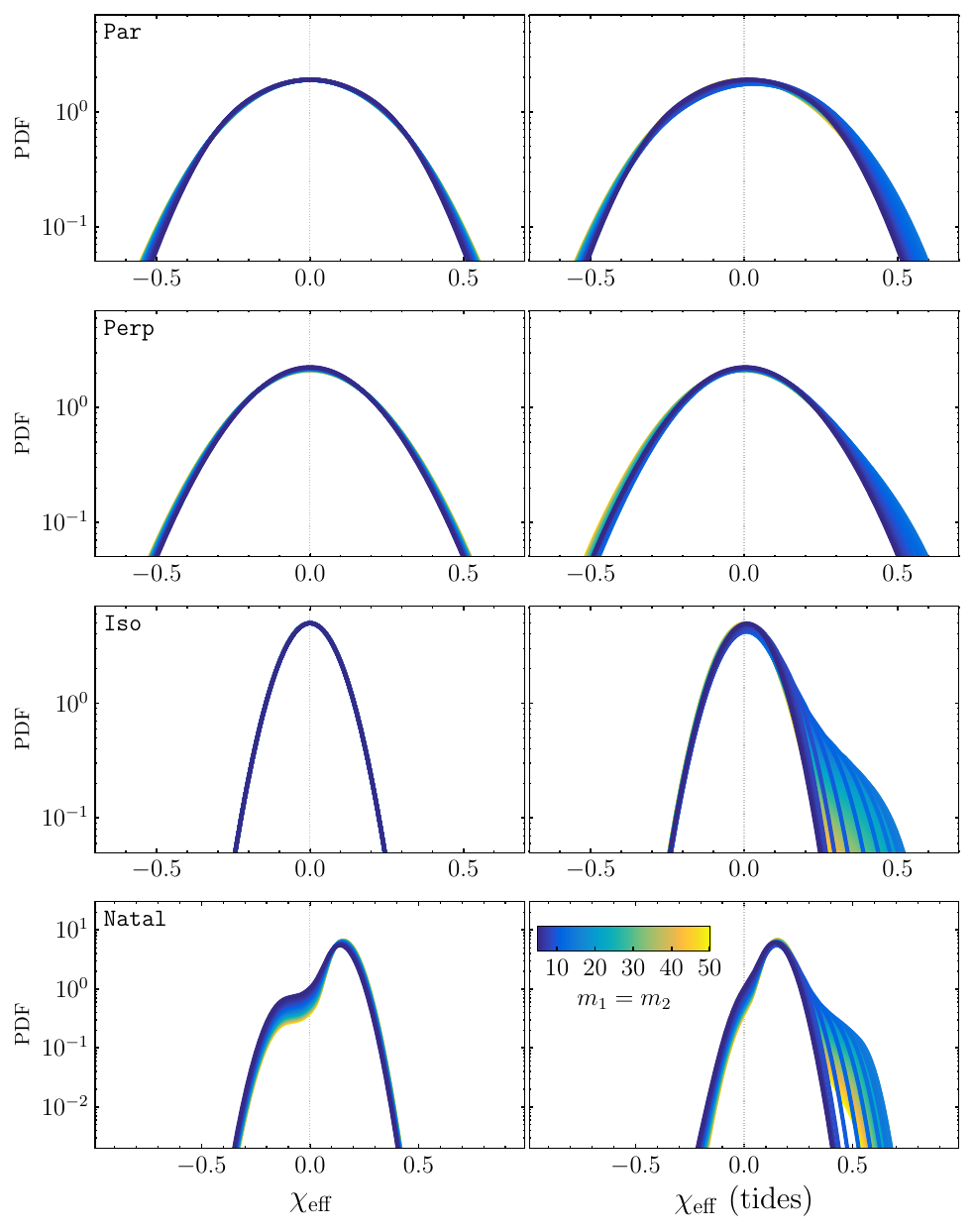}
    \caption{PDFs of \chieff based on the interacting model for pre-CC2 orbital separations and high kicks for all BH masses: $p(\chieff)$ assuming inefficient tides (left), and $p(\chieff)$ with tidal spin-up (right) for the \Spar, \Sperp, \Siso, and \Snat models (from top to bottom) for BH masses ranging from $m_1=m_2=5 \Msun$ (in blue) to $50 \Msun$ (in yellow).}
    \label{fig:chieff_MVK_250_interacting}
\end{figure}

\section{Population properties of merging black holes}
\label{sec:pop}

In this section, we  generate Monte Carlo populations of BBHs under the four spin-direction models discussed here: \Snat, \Siso, \Spar, and \Sperp. Our goal is to study how different spin-generation mechanisms affect the population properties observed by LVK, such as the distribution of spin--orbit misalignments and \chieff.
While the previous section focused on individual systems after CC and how their spin tilts vary as a function of natal kick magnitude, we now shift our attention to statistical properties across a population. Rather than simulating CC ourselves, we adopt analytic and Monte Carlo models based on the spin-generation mechanisms proposed in the CC literature and on the spin-kick correlations inferred from pulsar observations. We likewise do not attempt a full population synthesis with detailed stellar evolution modeling; a simplified prescription instead lets us isolate the impact of spin-direction assumptions on BBH spin-tilt distributions. This allows us to assess, in a controlled way, how different spin-kick couplings may imprint signatures on the spin orientations of merging BBHs observable through GWs.

\subsection{Population model}
Our models follow the evolution of a binary system comprising two stars with masses $M_1$ and $M_2$ initially separated by $a_{i1}$. After the first CC event (CC1), the star with mass $M_1$ transforms into a BH with mass $m_1$. Due to asymmetric ejecta, the BH experiences a recoil velocity $v_{k1}$, changing the semimajor axis to $a_{f1}$. At a later stage in the binary's evolution, when the separation is $a_{i2}$, the secondary star with mass $M_2$ undergoes CC (CC2), producing a BH with mass $m_2$. This event leads to mass loss and a recoil velocity $v_{k2}$,  changing the binary's semimajor axis to $a_{f2}$. Next, we will elaborate on the prescriptions we use for orbital separations, natal kick magnitudes, spin magnitudes, and tidal interactions.

\subsubsection{Semimajor axis}
\label{sec:semimajor_models}
We begin by sampling the initial binary separation, $a_{i1}$, from a uniform logarithmic distribution \citep{1924PTarO..25f...1O,Sana:2012px,2017ApJS..230...15M} in the range of $[10\Rsun, 10^4\Rsun]$.
Following the occurrence of CC1, the orbital separation changes to $a_{f1}$ (Equation~(\ref{eq:ae})), and CC2 takes place at a distance $a_{i2}$~\citep{1976IAUS...73...75P,Iben:1993yi,Ivanova:2012vx,Rocha:2024oqc}. Our goal is to understand how different distributions  of $a_{i2}$ affect the population properties, and we consider two extreme scenarios:

\subheading{Noninteracting} We assume that $a_{f1}=a_{i2}$, implying that the separation after CC1 and between CC2 remains unchanged. The evolution of the orbital separation is entirely governed by the natal kicks before the gravitational waves take over, driving the inspiral and reducing the separation until the BHs merge. However, in our synthetic population, a significant proportion of binaries fail to merge within the Hubble time due to the broad range of initial orbital separations sampled.

\subheading{Interacting} In reality, several physical processes such as stable mass transfer or common envelope can modify the binary separation between CC1 and CC2. Many binaries in various population synthesis models undergo at least one of these processes, leading to a significant reduction in orbital separation such that the BHs are close enough to merge via gravitational waves~\citep{Dominik:2012kk,Dominik:2013tma,Dominik:2014yma,Belczynski:2016obo,Mapelli:2019bnp,Santoliquido:2020axb,Gallegos-Garcia:2021hti,Marchant:2021hiv,vanSon:2021zpk,Stevenson:2022djs}. We account for their effects by assuming that the separation changes from $a_{f1}$ to $a_{i2}$ in such a way that after CC2, when the separation again changes to $a_{f2}$, it is small enough to allow a merger within the Hubble time, assuming zero eccentricity. This implies that all binaries that survived disruption during CC1 will merge after CC2. In doing so, our goal is to effectively mimic the impact of a common envelope phase by shrinking the binary orbit enough to enable a BBH merger.

\subsubsection{Natal kick magnitudes}
\label{sec:kick_models}
At present, there is no clear consensus on the magnitude of BH kicks. The inferences drawn from the orbital parameters, space velocities, and locations of individual binaries within the Galaxy are inconclusive. While some binary systems, such as Cygnus X-1 and GRS 1915+105, do not require natal BH kicks or suggest only a small kick \citep{Nelemans:1999ux, Wong:2010ak, Banagiri:2022ayy}, others, like MAXI J1305-704 and XTE J1118+480, indicate large kicks  \citep{1995MNRAS.277L..35B, 2000IAUC.7389....2R, Mirabel:2001ay, Willems:2004kk, Gualandris:2004tv, Fragos:2008hg, Kimball:2022xbp}. Population synthesis studies \citep{Mapelli:2018wys, Giacobbo:2018etu, Chruslinska:2018ylm, Baibhav:2019gxm} indicate that very low kicks are needed for merging compact objects to explain the observed merger rate density from LIGO-Virgo data. \citet{Janka:2013hfa} also argued that if kicks are driven by neutrinos, they will be suppressed for massive BHs; however, if kicks are driven by explosion asymmetries, they can compete with the high kicks experienced by neutron stars, reaching several hundred kilometers per second. To accommodate the two extreme scenarios, we explore two extreme kick prescriptions:

\subheading{High kicks} All BHs, regardless of their other properties, are given a kick randomly drawn from a Maxwellian distribution with a root mean square (rms) value of $v_k=250\, \kms$, similar to those experienced by neutron stars~\citep{Hobbs:2005yx, Faucher-Giguere:2005dxp}.

\subheading{Reduced kicks} In this model, we assume that while lighter BHs can receive large kicks, heavier BHs will receive lower kicks to account for the effects of fallback on the formation of BHs~\citep{Fryer:2011cx,Giacobbo:2019fmo}. We assume that the more massive the BH, the larger the amount of fallback it experiences, which in turn reduces its natal kick. In this model, we impose that the magnitude of the natal kick decreases linearly from $265\ \kms$ for a $1.4\ \Msun$ neutron star to $0\ \kms$ for a $20\ \Msun$ BH.  In addition, we assume that BHs with masses greater than $20\Msun$ form via direct collapse of a massive star, and do not receive any natal kicks or mass loss during CC.

\subsubsection{Spin magnitudes}
 In Section~\ref{sec:LS}, we establish models that describe the direction of the spin. However, determining the magnitude of the spin is significantly more challenging. This is because the mechanisms that influence the magnitude of the spin are not yet well understood. In this article, we consider two models for the spin magnitude.

\subheading{Spins triggered by natal kicks} In the \Spar and \Sperp models, the BH spin is generated by the same process that imparts the natal kick, so the two are expected to scale together. The CC simulations of \citet{Muller:2018utr} find precisely such a correlation between kick and spin magnitude (their Figure~12), and the same scaling follows from the off-center explosions of \citet{Spruit:1998sg}. We therefore adopt the simplest relation consistent with both,
\be
\chi=\hat{\chi} \, v_{\rm kick}\,,
\ee
where $\hat{\chi}$ is a constant of proportionality. In this study, we have set the free parameter $\hat{\chi}=0.2/100$ for illustrative purposes, such that a kick with a magnitude of $100 \kms$ would produce a BH with  spin magnitude of $0.2$. We will employ this prescription for models \Spar and \Sperp.

\subheading{Maxwellian spins} In addition to kicks, BHs can also be spun through various stochastic processes that increase the spin via a random walk, potentially leading to isotropic spins. When the spin is built up through many independent, random increments, the resulting distribution of spin magnitudes is Maxwellian. An example of this is the spinning of stellar cores due to internal gravity waves originating from the convective layer, as suggested by \citet{2014ApJ...796...17F, 2015ApJ...810..101F, Ma:2019cpr, McNeill:2020hbp}.
In order to model the \Siso scenario, we will use a Maxwellian distribution with an rms value of $0.2$ for the spin magnitudes. We will also use the same distribution for the \Snat model for simplicity.

\subsubsection{Spin-tilt evolution and tidal alignment}

After CC1, the binary system undergoes changes in its orientation and the BH spin. The post-CC1 spin--orbit misalignment  $\theta_{{LS}_1}$ will depend on the spin-direction models discussed in Section~\ref{sec:LS}.  The subsequent CC, CC2, will tilt the orbit by $\theta_{{LL}_{i2}}$, and impart spin to the second-born BH for models \Spar, \Sperp, and \Siso.  The tilt of the first-born BH with respect to the final orbital angular momentum will change again as
\be\label{eq:costheta1}
\cos \theta_1 = \cos\theta_{{LS}_1} \cos\theta_{{LL}_{i2}} + \sin\theta_{{LS}_1}\sin\theta_{{LL}_{i2}}\sin\phi_{S1},
\ee
where $\cos\theta_{{LS}_1}$ is spin--orbit misalignment post-CC1, while $\phi_{S1}$ is sampled uniformly between $[0, 2\pi]$.
In addition to previously discussed spin-generation mechanisms, tidal interactions can also contribute to the spin of the BH. Tides can synchronize stellar spins with the binary orbit, potentially resulting in BH spin parallel to the orbital angular momentum after the CC \citep{Hurley:2002rf}. While a detailed treatment of tidal theory in massive stars is beyond the scope of this paper, we consider two possibilities: efficient tidal spin-up of the secondary star along the orbital angular momentum, or completely inefficient tides \citep{Eggleton:2001fs, 2006epbm.book.....E}.

\subheading{No tides} Since there are no tidal effects, the tilt for the second-born BH can be directly calculated as specified in Section~\ref{sec:LS}.

\subheading{Tides} In this scenario, we will focus on the tidal effects on the secondary star between the two CC events in the binary evolution. During this phase, if tidal interactions are efficient, they tend to spin the star (but not the BH) with the orbital angular momentum. To model this effect, we will use fits of the dimensionless spin of the second-born BH by \citet{Bavera:2021evk} as a function of binary orbital period, obtained by fitting detailed MESA simulations that accurately tracked the evolution and spin-up of close BH--Wolf--Rayet systems. The second-born BH will have two sources of spin: spin along the pre-CC2 orbital angular momentum $\boldsymbol{\chi_t}$ calculated using fits by \citet{Bavera:2021evk}; and spin $\boldsymbol{\chi_{2,0}}$ due to different processes discussed in Section \ref{sec:LS}, with a total spin of $\boldsymbol{\chi_{2}}=\boldsymbol{\chi_t}+\boldsymbol{\chi_{2,0}}$, and spherical angles given by $(\theta_{{LS}_2},\phi_{{LS}_2})$ with respect to the pre-CC2 orbital angular momentum. The misalignment with respect to the final orbital angular momentum is given by
\be\label{eq:costheta2}
\cos \theta_2 = \cos\theta_{{LS}_2} \cos\theta_{{LL}_{i2}} + \sin\theta_{{LS}_2}\sin\theta_{{LL}_{i2}}\sin\phi_{{LS}_2},
\ee
where $\theta_{{LL}_{i2}}$ is the tilt angle between the final and pre-CC2 orbital angular momenta.

\begin{figure*}[t]
   \includegraphics[width=\textwidth]{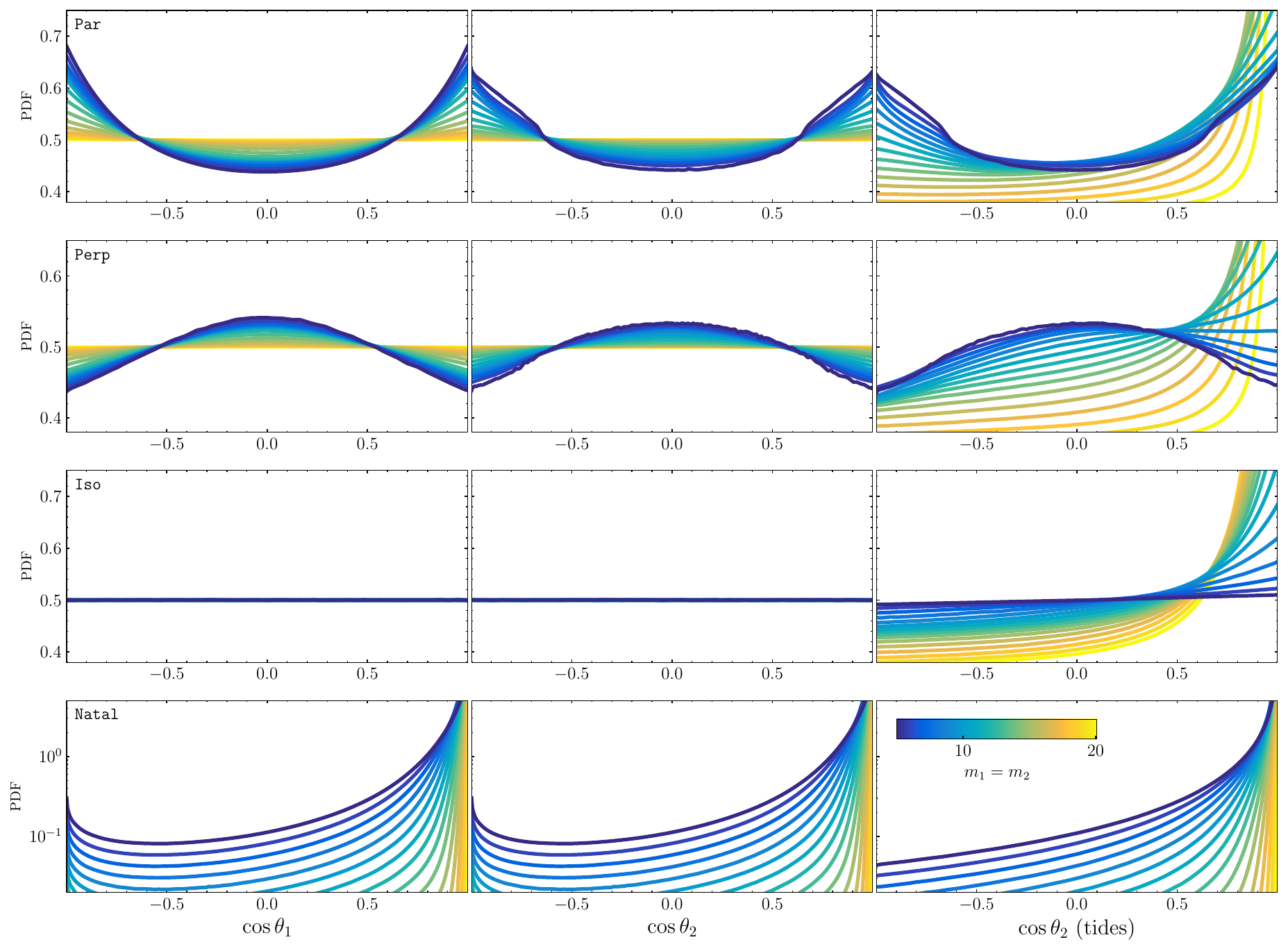}
    \caption{PDFs of spin--orbit tilts based on the noninteracting model for pre-CC2 orbital separations and reduced kicks for higher BH masses: $p(\cos\theta_1)$ (left), $p(\cos\theta_2)$ assuming inefficient tides (middle), and $p(\cos\theta_2)$ with tidal spin-up (right) for the \Spar, \Sperp, \Siso, and \Snat models (from top to bottom) for BH masses ranging from $m_1=m_2=5 \Msun$ (in blue) to $20 \Msun$ (in yellow).}
    \label{fig:costheta_M_noninteracting}
\end{figure*}

\begin{figure*}[t]
   \includegraphics[width=\textwidth]{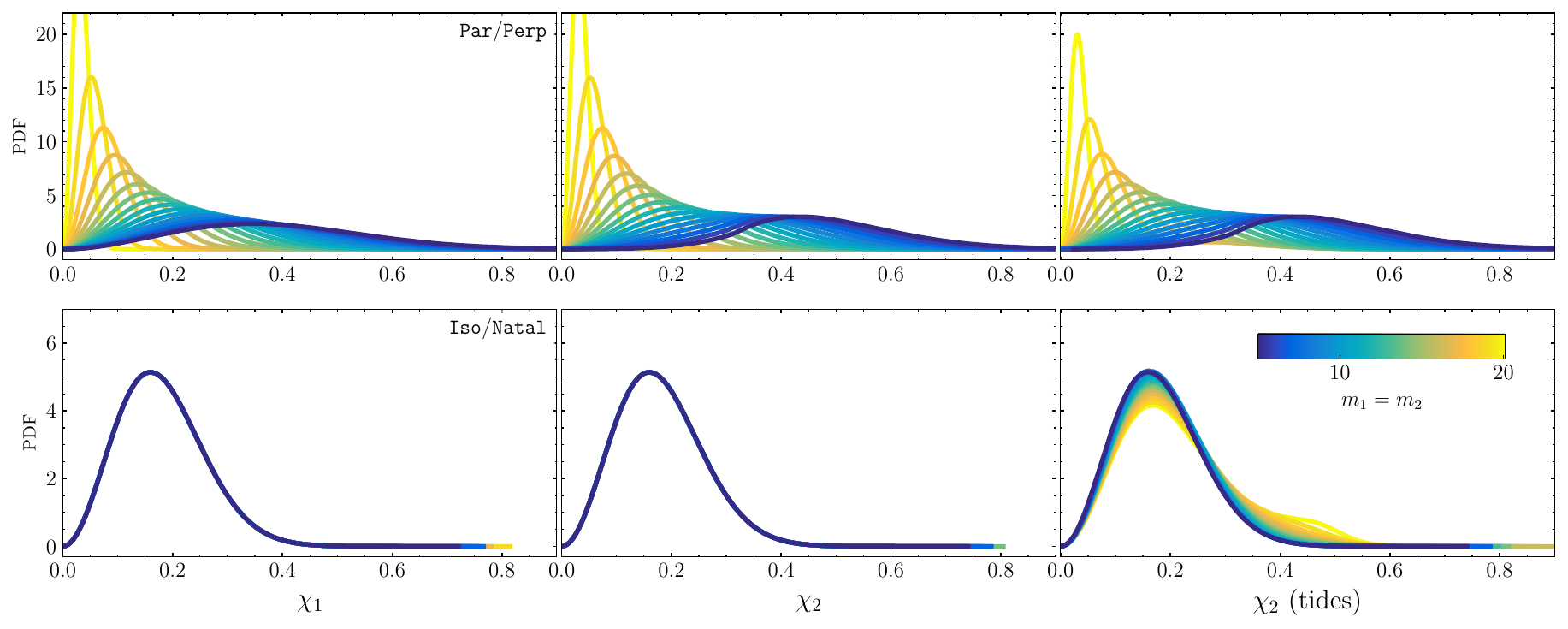}
    \caption{PDFs of spin magnitudes based on the noninteracting model for pre-CC2 orbital separations and reduced kicks for higher BH masses: $p(\chi_1)$ (left), $p(\chi_2)$ assuming inefficient tides (middle), and $p(\chi_2)$ with tidal spin-up (right) for BH masses ranging from $m_1=m_2=5 \Msun$ (in blue) to $20 \Msun$ (in yellow). The top panel shows the distribution for \Spar and \Sperp where spin magnitudes were assumed to be proportional to the natal kick, while the bottom panel shows the distributions for \Siso and \Snat models where spin magnitudes are drawn from a Maxwellian distribution.}
    \label{fig:chi_M_noninteracting}
\end{figure*}

\begin{figure*}[t]
   \includegraphics[width=\textwidth]{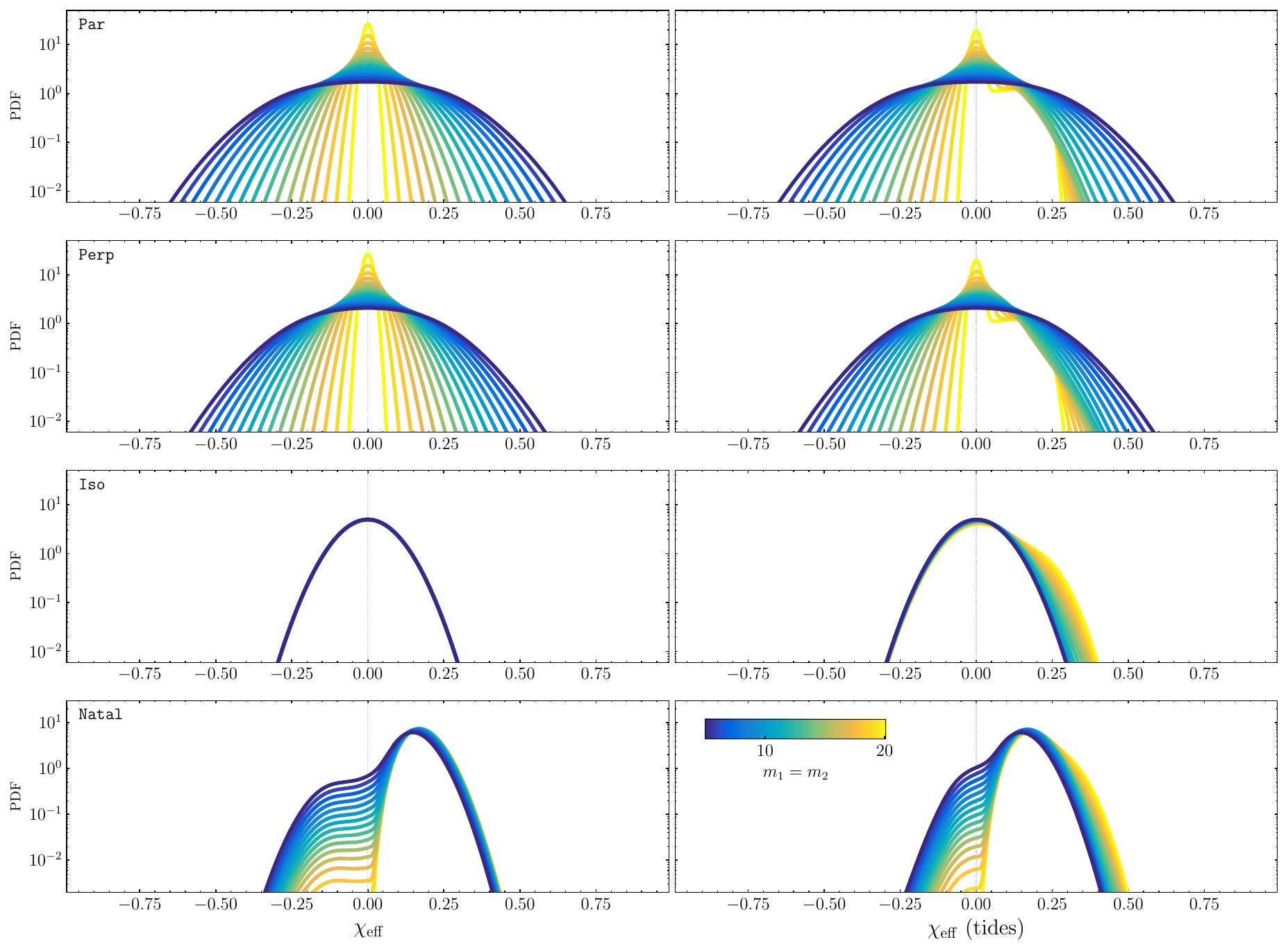}
    \caption{PDFs of \chieff based on the noninteracting model for pre-CC2 orbital separations and reduced kicks for higher BH masses: $p(\chieff)$ assuming inefficient tides (left), and $p(\chieff)$ with tidal spin-up (right) for the \Spar, \Sperp, \Siso, and \Snat models (from top to bottom) for BH masses ranging from $m_1=m_2=5 \Msun$ (in blue) to $20 \Msun$ (in yellow).}
    \label{fig:chieff_M_noninteracting}
\end{figure*}

\begin{figure*}[t]
   \includegraphics[width=\textwidth]{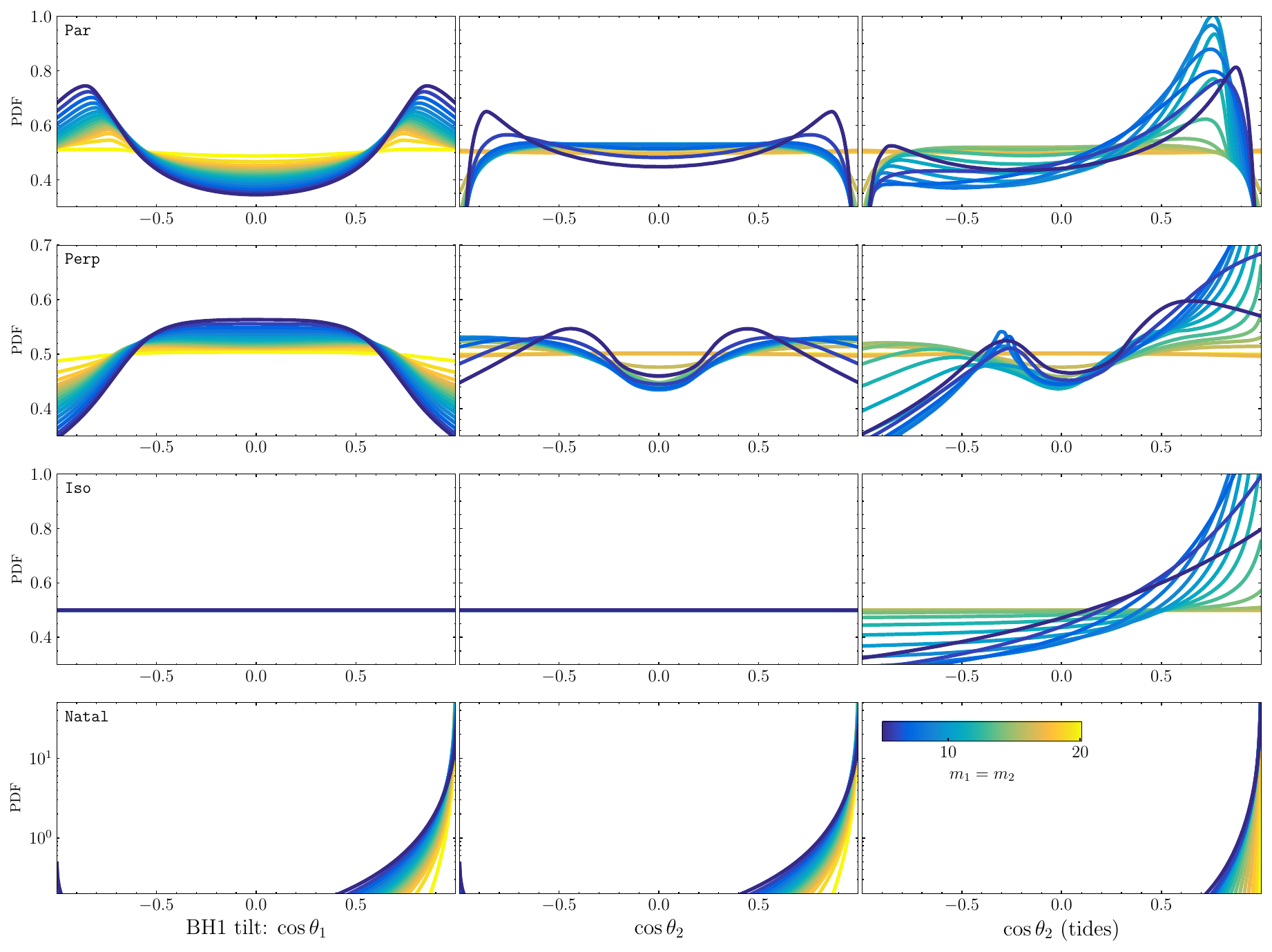}
    \caption{PDFs of spin--orbit tilts based on the interacting model for pre-CC2 orbital separations and reduced kicks for higher BH masses: $p(\cos\theta_1)$ (left), $p(\cos\theta_2)$ assuming inefficient tides (middle), and $p(\cos\theta_2)$ with tidal spin-up (right) for the \Spar, \Sperp, \Siso, and \Snat models (from top to bottom) for BH masses ranging from $m_1=m_2=5 \Msun$ (in blue) to $20 \Msun$ (in yellow).}
    \label{fig:costheta_M_interacting}
\end{figure*}

\subsection{Distribution of GW observables}
In the previous section, we established a synthetic population, and the next step is to examine how the distribution of various GW spin observables changes in response to the processes that drive BHs to spin. Specifically, we will study how different prescriptions for factors such as orbital separation, tides, spin magnitudes, and natal kicks will impact spin magnitudes, spin--orbit tilts, and effective inspiral spin,
 \be
 \chieff= \frac{m_1 \chi_1 \cos\theta_1 + m_2 \chi_2 \cos\theta_2 }{m_1+m_2}.
 \ee

\subsubsection{High kicks}

Let us consider the extreme case that BHs receive a natal kick  drawn from a Maxwellian distribution with an rms value of $v_{k}=250\ \kms$.

\subheading{Noninteracting binaries} In this scenario, the binary separation remains unchanged before CC2 and after CC1. We present the spin--orbit tilt distribution in Figure~\ref{fig:costheta_MVK_250_noninteracting}, covering a range of BH masses from $m_1=m_2=5 \Msun$ (shown in blue) to $50 \Msun$ (shown in yellow). The four panels from top to bottom correspond to models \Spar, \Sperp, \Siso, and \Snat. The first and second columns display the distribution of $\cos\theta_1$ and $\cos\theta_2$ assuming inefficient tides, while the third column shows the distribution of $\cos\theta_2$ if tides are allowed.
We find a slight preference for aligned and anti-aligned configurations in $\cos\theta_{1,2}$ for the \Spar model, whereas the \Sperp model prefers in-plane spins. This is because in-plane kicks are disfavored, and therefore, if the spins are perpendicular (parallel) to the kicks, they will disfavor spins along (perpendicular to) the orbital angular momentum. The \Siso model shows a uniform distribution of $\cos\theta_{1,2}$, while the \Snat model displays mostly aligned BHs with tails extending to larger misalignments. It is worth noting that lighter BHs tend to have larger misalignments in the \Snat model and greater deviations from isotropy in the \Spar and \Sperp models.

This behavior can be understood as follows: the misalignments in the \Snat model and deviations from isotropy in the \Spar and \Sperp models depend on $v_k/v_0 \propto M_i^{-1/2}$, where $M_i$ is the total binary mass pre-CC. For binaries that receive the same kick and have the same orbital separation, binaries with a larger total binary mass pre-CC have smaller $u_k$ and hence smaller misalignments and fewer binary disruptions.
Additionally, in the third column of Figure~\ref{fig:costheta_MVK_250_noninteracting}, we also plot the distributions of $\cos\theta_2$ if tides are allowed, and some binaries get a larger spin component along the orbital angular momentum. This spin-up is larger for massive binaries, and in all four models discussed, they tend to be more aligned when tides are allowed.

In this scenario, all BHs have the same kick distribution, resulting in the same spin magnitude distribution for the \Spar and \Sperp models, regardless of their masses. On the other hand, the spin magnitudes for the \Siso and \Snat models follow a Maxwellian distribution. The distribution of $\chieff$ is presented in Figure~\ref{fig:chieff_MVK_250_noninteracting}. For \Siso, \Sperp, and \Spar models, the distribution of $\chieff$ is symmetric around $\chieff=0$, while for \Snat, it is mostly positive with a tail toward negative values~\citep{Gerosa:2018wbw, Callister:2020vyz, Banerjee:2023ycw}. Since the spin magnitude distribution for both \Spar and \Sperp models is the same, this results in similar $\chieff$ distributions as well. Moreover, when tides are allowed to spin up the binary, we observe a larger effect in higher-mass BH binaries, where their spins tend to be more aligned with the orbital angular momentum and have a positive $\chieff$ distribution.

\subheading{Interacting binaries} We next consider the interacting prescription defined in Section~\ref{sec:semimajor_models}. In Figures~\ref{fig:costheta_MVK_250_interacting} and~\ref{fig:chieff_MVK_250_interacting}, we illustrate the spin--orbit tilt distribution and $\chieff$ distributions for a range of BH masses from $m_1=m_2=5 \Msun$ (shown in blue) to $50 \Msun$ (shown in yellow). In Figure~\ref{fig:costheta_MVK_250_interacting}, we show the four panels from top to bottom corresponding to models \Spar, \Sperp, \Siso, and \Snat. The first and second columns of each panel display $\cos\theta_1$ and $\cos\theta_2$, assuming inefficient tides, while the third column shows the distribution of $\cos\theta_2$ if tides are allowed. We find that the spin--orbit misalignment distribution for \Spar has peak tilts from aligned/anti-aligned by around $40^\circ$ for the first-born BH and $20^\circ-30^\circ$ for the second-born BH. On the other hand, the distribution of $\cos\theta_2$ has less pronounced features because CC2 goes off at much smaller orbital separations. The perpendicular-to-kick spin-direction model, \Sperp, does not prefer aligned or anti-aligned configurations. Once again, tides break the symmetry around $\cos\theta_2=0$. The distributions we observe are markedly distinct from those in the noninteracting case we presented earlier. This is because the effect of natal kicks is more prominent when the binary's initial separation is large (assuming the binary remains bound after the CC). In the previous case, if the binary's initial separation was large, the probability of its orbit shrinking enough to merge within the Hubble time was low. Therefore, binaries with large initial separations were excluded from the distributions of merging binaries. However, in this case, all binaries that survive the first CC merge. After CC1, the binary system undergoes changes in
its orientation and spin, which means that even binaries with very large orbital separations after CC1 are included in the distributions shown in Figures~\ref{fig:costheta_MVK_250_interacting} and~\ref{fig:chieff_MVK_250_interacting}.

\subsubsection{Reduced kicks}

We now apply the reduced-kick prescription defined in Section~\ref{sec:kick_models}, in which fallback suppresses the kicks of more massive BHs.

\subheading{Noninteracting binaries} We present the spin--orbit tilt distribution in Figure~\ref{fig:costheta_M_noninteracting}, covering a range of BH masses from $m_1=m_2=5 \Msun$ (shown in blue) to $20 \Msun$ (shown in yellow), assuming that the binary separation remains unchanged between CC1 and CC2. The four panels from top to bottom correspond to models \Spar, \Sperp, \Siso, and \Snat, with the first and second columns displaying $\cos\theta_1$ and $\cos\theta_2$ assuming inefficient tides, while the third column shows the distribution of $\cos\theta_2$ if tides are allowed. In Figure~\ref{fig:costheta_MVK_250_noninteracting}, when all BHs received high kicks, the deviation from isotropic spin--orbit tilts was more pronounced in lighter BHs compared to heavier ones. However, this effect gets amplified if we assume that lighter BHs receive more kicks  compared to heavier ones. This can be observed in Figure~\ref{fig:costheta_M_noninteracting}, where heavy BHs always appear to be isotropic while lighter ones have a slight preference for aligned/anti-aligned spins for \Spar and in-plane spins for \Sperp. Additionally, lighter $5 \Msun$ BHs have longer distribution tails away from $\cos\theta_{1,2}=1$, while heavier BHs are very strictly aligned in the \Snat model, for the same reason.

 Since the spin magnitudes in the \Spar and \Sperp models are proportional to the kick magnitude, and we assume that BHs of different masses receive different natal kicks, BHs of different masses should have different spin magnitude distributions. This can be seen in the top panel of Figure~\ref{fig:chi_M_noninteracting}, where we plot the distributions of $\chi_1$, $\chi_2$, and $\chi_2$ with spin up by tides for the \Spar and \Sperp models. We observe that for higher masses, the spin magnitudes are small, while for lighter masses, the spin magnitudes are large. This also translates to a correlation between \chieff and mass, as shown in Figure~\ref{fig:chieff_M_noninteracting} for both \Spar (top panel) and \Sperp (second panel). We find that as we increase the component masses, the \chieff distribution becomes narrower. On the other hand, the spin magnitudes for the \Siso and \Snat models follow a Maxwellian distribution, as shown in Figure~\ref{fig:chi_M_noninteracting}. We also plot the \chieff distribution for these two models in Figure~\ref{fig:chieff_M_noninteracting}. The \chieff distribution in the \Siso model is symmetric about $\chieff=0$, while the \chieff distribution in the \Snat model is mostly positive with a tail toward negative values. In the \Snat model, lighter BHs receive larger kicks and have a larger effect of spin--orbit tilts, resulting in larger negative $\chieff$ tails. Furthermore, when tides are allowed to spin up the binary, we observe a larger effect in higher-mass BH binaries, where their spins tend to be more aligned with the orbital angular momentum and more binaries have positive \chieff values.

\subheading{Interacting binaries} For interacting binaries under the reduced-kick prescription, Figure~\ref{fig:costheta_M_interacting} illustrates the spin--orbit tilt distributions for a range of BH masses, from $m_1=m_2=5 \Msun$ (shown in blue) to $20 \Msun$ (shown in yellow). Figure~\ref{fig:costheta_M_interacting} displays four panels from top to bottom corresponding to models \Spar, \Sperp, \Siso, and \Snat. The first and second columns of each panel display $\cos\theta_1$ and $\cos\theta_2$, respectively, assuming inefficient tides, while the third column shows the distribution of $\cos\theta_2$ if tides are allowed. The peaks and features of the distributions are qualitatively similar to those shown in Figure~\ref{fig:costheta_MVK_250_noninteracting}, where all BHs received high kicks. However, in this case, heavier BHs receive smaller kicks, resulting in less pronounced peaks and closer-to-isotropic distributions for $\Spar$ and $\Sperp$. We also observe that the inclusion of tides can impact the distribution's symmetry, with the peak at negative $\cos\theta_2$ being suppressed and the peak at positive $\cos\theta_2$ being amplified. For example, the peak for $10\Msun$ BHs lies at $\theta_2=40^\circ$, where the bin at this peak contains twice as many binaries as if $\cos\theta_2$ were uniform. Finally, we note that this distribution shares qualitative similarities with GWTC-3 population analyses that reported possible structure around $\theta\simeq60^\circ$, rather than a peak at alignment~\citep{Vitale:2022dpa, Edelman:2022ydv, Callister:2023tgi}.

\section{Comparison with LVK observations}
\label{sec:LVK}

In this section, we compare the primary spin-tilt distributions predicted by the four models with observations of BBH mergers in GWTC-5.0~\citep{LIGOScientific:2026sit}. Spin orientations are commonly used to distinguish between isolated binary evolution and dynamical assembly~\citep{Vitale:2015tea, Rodriguez:2016vmx, Stevenson:2017dlk, Talbot:2017yur, Farr:2017uvj, Gerosa:2018wbw}. Here we ask a complementary question: can measurements of the primary spin tilt distinguish among different spin-generation mechanisms within isolated binary evolution? Conditional on each model, we also infer the normalized natal kick magnitudes favored by the data.

\subsection{Data and assumptions}
\label{sec:LVK_method}

We use the publicly released parameter-estimation (PE) posterior samples for 257 BBH mergers selected from GWTC-5.0~\citep{LIGOScientific:2026sit}. Candidates classified as potentially containing a neutron star are excluded from the BBH sample because the spin-generation models considered here apply specifically to BH formation. We analyze GW230529 separately, applying these models only to its primary compact object under the interpretation that the primary is a BH. This analysis is intended as a deliberately simple, first-order calculation that isolates how the assumed origin of the BH spin changes the qualitative kick inference, rather than as a complete reconstruction of the binary population. A complete population model analysis is out of scope, given the focus of our study.

We restrict the tilt-dependent likelihood to the primary tilt, $\cos\theta_1$, because the secondary tilt is weakly constrained for most events. We also assume that the primary BH is the first-born BH, allowing us to identify $\cos\theta_1$ with the spin--orbit misalignment produced at CC1. Both choices are modeling approximations: some events contain information about correlated primary and secondary tilts, and mass transfer can reverse the component-mass ordering. We discuss their implications in Section~\ref{sec:LVK_sys}.

Each binary undergoes two CC events, with normalized kick magnitudes $u_{k1}=v_{k1}/v_{{\rm orb},1}$ and $u_{k2}=v_{k2}/v_{{\rm orb},2}$. Unless stated otherwise, we set $u_{k2}=0$ and infer $u_{k1}$. This approximation is motivated by the smaller separation and therefore larger orbital velocity expected before CC2, which reduces the normalized second kick for a fixed physical kick magnitude. In this limit, CC2 does not tilt the orbital plane and, under the first-born-primary assumption, $\cos\theta_1=\cos\theta_{{LS}_1}$. We therefore treat CC1 as the dominant source of the primary spin--orbit misalignment. In the same limit, the \Siso, \Spar, and \Sperp models approach a uniform distribution in $\cos\theta_2$, whereas \Snat predicts $\cos\theta_2=1$. Because current observations provide little information about $\cos\theta_2$, these different predictions contribute little to the present model comparison. The limitations of the $u_{k2}=0$ approximation are discussed in Section~\ref{sec:LVK_sys}.

Throughout this section, we assume circular precollapse orbits and set $\beta=1$, corresponding to negligible mass loss during each collapse. The no-mass-loss limit is most applicable to BHs formed with little ejecta and allows us to isolate the effects of natal kicks on the spin tilts. As shown in Section~\ref{sec:LS}, varying $\beta$ changes the disruption boundaries, $\sqrt{2\beta}\pm1$, but preserves the qualitative features of the predicted tilt distributions. Tidal interactions provide a physical motivation for orbital circularization before CC2; circularity before CC1 is an additional assumption of our analysis. At $\beta=1$, no binary remains bound when the normalized kick exceeds $1+\sqrt{2}$ [Equation~(\ref{eq:cosLLRange})]. We therefore restrict $u_{k1}$ and $u_{k2}$ to the interval $[0,\,1+\sqrt{2}]$.

We adopt a prior on $u_{k1}$ proportional to the probability that the binary remains bound after CC1. For an isotropically oriented kick and $\beta=1$, the survival probability follows from the disruption condition derived in Section~\ref{sec:LS}. After normalization over $u_{k1}\in[0,\,1+\sqrt{2}]$, the prior is constant, $p_{\rm prior}(u_{k1})=1/C$, for $0\leq u_{k1}<\sqrt{2}-1$. Once disruption becomes possible, it decreases as
\be\label{eq:psurv}
p_{\rm prior}(u_{k1})
=
\frac{2-(1-u_{k1})^2}{4C u_{k1}},
\qquad
\sqrt{2}-1\leq u_{k1}\leq\sqrt{2}+1.
\ee
The normalization constant is
\be
C=\frac{1}{\sqrt{2}}
+\frac{1}{4}\ln\!\left(3+2\sqrt{2}\right).
\ee
We use the same functional form for $u_{k2}$ when the second kick is allowed to vary.

\subsection{Inference on individual events}
\label{sec:LVK_events}

For each event and spin-direction model $M$, the predicted primary-tilt distribution is $p_M(\cos\theta_1\mid u_{k1})$. We compare it with the marginalized PE posterior $p_{\rm PE}(\cos\theta_1\mid d)$. For the PE samples used here, the corresponding prior is uniform, $\pi_{\rm PE}(\cos\theta_1)=1/2$, so this constant can be absorbed into the proportionality. The posterior for the normalized kick is therefore
\be\label{eq:event_like}
\begin{aligned}
p_M(u_{k1}\mid d)
&\propto
p_{\rm prior}(u_{k1})
\int \dd\cos\theta_1\\[-2pt]
&\quad\times
p_{\rm PE}(\cos\theta_1\mid d)\,
p_M(\cos\theta_1\mid u_{k1}).
\end{aligned}
\ee
Here, $p_M(\cos\theta_1\mid u_{k1})$ is evaluated among binaries that remain bound after CC1, with $\beta=1$ and $u_{k2}=0$. It corresponds to the distribution of $\cos\thetaLS$ derived in Section~\ref{sec:LS}, except for \Snat, for which $\cos\theta_1=\cos\thetaLL$.

The normalization of Equation~(\ref{eq:event_like}) gives the evidence $\mathcal{Z}_M$, which we evaluate using \code{dynesty}~\citep{Speagle:2019ivv}. Throughout this section, we report log Bayes factors relative to \Snat, $\ln{\rm BF}_{M/{\rm Natal}}=\ln\mathcal{Z}_M-\ln\mathcal{Z}_{\rm Natal}$; positive values favor model $M$, while negative values favor \Snat. This individual event comparison is conditional on the observed events and does not include catalog-level selection effects.

\begin{figure}[t]
   \includegraphics[width=\columnwidth]{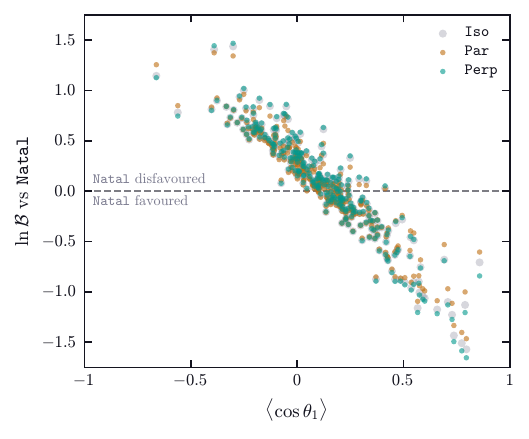}
    \caption{Per-event log Bayes factors relative to \Snat as a function of the posterior mean primary tilt, $\langle\cos\theta_1\rangle$, for the 257 BBH mergers. Grey, orange, and teal markers show \Siso, \Spar, and \Sperp, respectively. Positive values disfavor spin inheritance, whereas negative values favor it. The three alternatives follow nearly the same trend because the \Spar and \Sperp event-level evidences remain close to \Siso. Events with predominantly aligned primary spins favor \Snat, while those with support for retrograde spins favor the alternatives.}
    \label{fig:lnbf_tilt}
\end{figure}

Figure~\ref{fig:lnbf_tilt} shows a strong relation between the measured primary tilt and the relative support for spin inheritance. Across the three alternative models, the log Bayes factors relative to \Snat range from approximately $-1.6$ to $1.5$ and are therefore modest for every individual event. At $u_{k1}=0$, \Snat predicts perfect alignment; increasing $u_{k1}$ broadens the distribution, and the full range of spin tilts becomes accessible only for $u_{k1}\geq1$ [Equation~(\ref{eq:cosLLRange})]. In contrast, \Spar and \Sperp predict isotropic primary tilts for $u_{k1}<\sqrt{2}-1$, because every kick direction leaves the binary bound in this regime. Their characteristic peaks at $\cos\thetaLS=\pm1/u_{k1}$ and $\cos\thetaLS=\pm\sqrt{u_{k1}^2-1}/u_{k1}$, given by Equations~(\ref{eq:ParPeak}) and~(\ref{eq:PerpPeak}), appear only for $u_{k1}>1$, where the survivability prior decreases with increasing kick magnitude. Consequently, the largest event-level differences occur between \Snat and the other three models.

\begin{figure*}[t]
   \includegraphics[width=\textwidth]{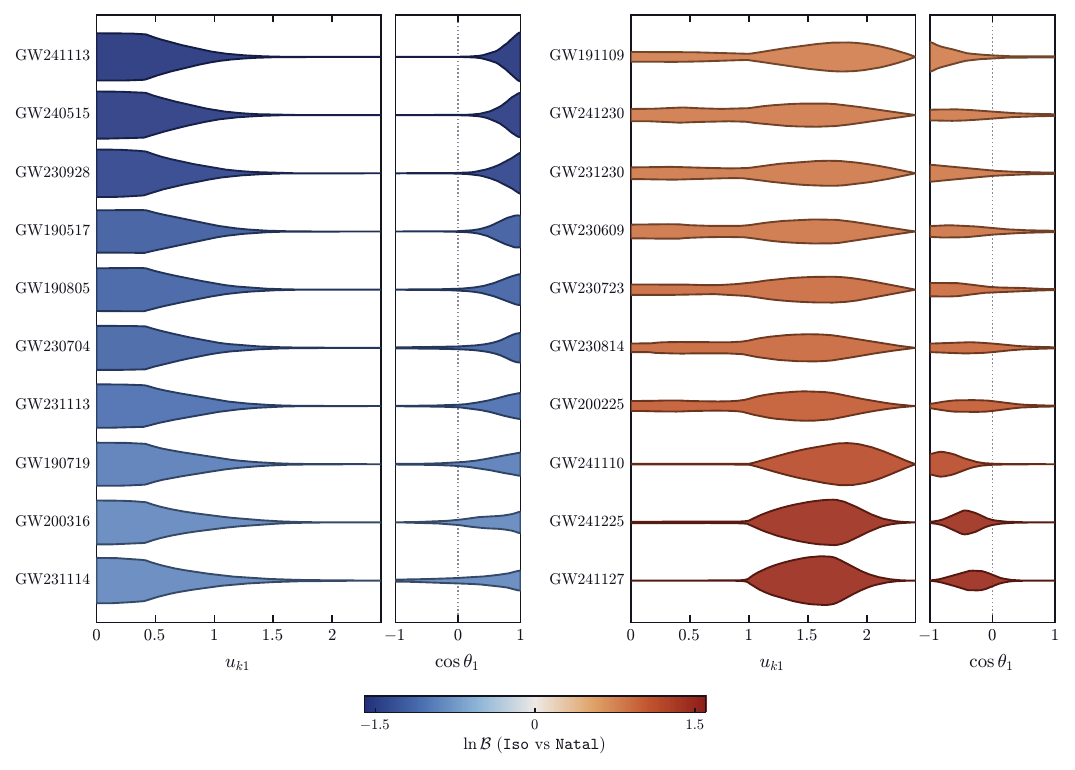}
    \caption{Natal model posteriors for $u_{k1}$ and the corresponding PE posteriors for $\cos\theta_1$ for the 10 events with the smallest inferred kicks (left) and the 10 with the largest inferred kicks (right). Color denotes $\ln{\rm BF}_{\rm Iso/Natal}$, and rows are ordered from greater to smaller support for \Snat.}
    \label{fig:event_kicks}
\end{figure*}

\textit{Events consistent with spin inheritance.} Ten events yield $\ln{\rm BF}_{\rm Iso/Natal}<-1$: GW240515\_005301, GW241113\_163507, GW230928\_215827, GW190412, GW190517, GW230704, GW170729, GW190805, GW231113\_122623, and GW231118\_005626. Their primary-tilt posteriors place substantial support near $\cos\theta_1=1$. Under \Snat, such tilts arise naturally from an initially aligned spin followed by a small CC1 kick, whereas \Siso assigns equal probability density across the full range of $\cos\theta_1$. Their \Snat kick posteriors consequently retain substantial support throughout the low-kick regime, as shown in the left half of Figure~\ref{fig:event_kicks}.

\textit{Events challenging spin inheritance.} Four events yield $\ln{\rm BF}_{\rm Iso/Natal}>1$: GW241127\_061008, GW241225\_082815, GW241110\_124123, and GW200225. Their primary-tilt posteriors place substantial support at large misalignments, including configurations with $\cos\theta_1<0$.

Under \Snat with $u_{k2}=0$, the primary tilt is produced entirely by reorientation of the orbital plane at CC1. Equation~(\ref{eq:cosLLRange}) requires $\cos\theta_1>\sqrt{1-u_{k1}^2}>0$ for $u_{k1}<1$; a kick smaller than the precollapse orbital velocity therefore cannot produce a retrograde primary. Consequently, support at $\cos\theta_1<0$ shifts the kick posterior toward $u_{k1}\gtrsim1$, where binary survival becomes less likely.

This behavior is clearest for GW241127\_061008, GW241225\_082815, and GW241110\_124123, whose $90\%$ credible intervals under \Snat exclude kicks smaller than the precollapse orbital velocity. The \Siso, \Spar, and \Sperp models instead permit retrograde spins without requiring a comparably large reorientation of the orbital plane. At current precision, however, the \Spar and \Sperp evidences remain close to \Siso, so individual events do not distinguish among these alternative spin-generation models.

Figure~\ref{fig:event_kicks} collects the \Snat kick posteriors for the events at the two extremes of the inferred kick distribution. For the most strongly misaligned events, the upper tails approach the disruption boundary $u_{k1}=1+\sqrt{2}$ and are therefore shaped by the requirement that the binary remain bound. These posteriors constrain the kicks among surviving binaries; they should not be interpreted as excluding larger kicks in the complete progenitor population.

\begin{figure*}[t]
   \includegraphics[width=\textwidth]{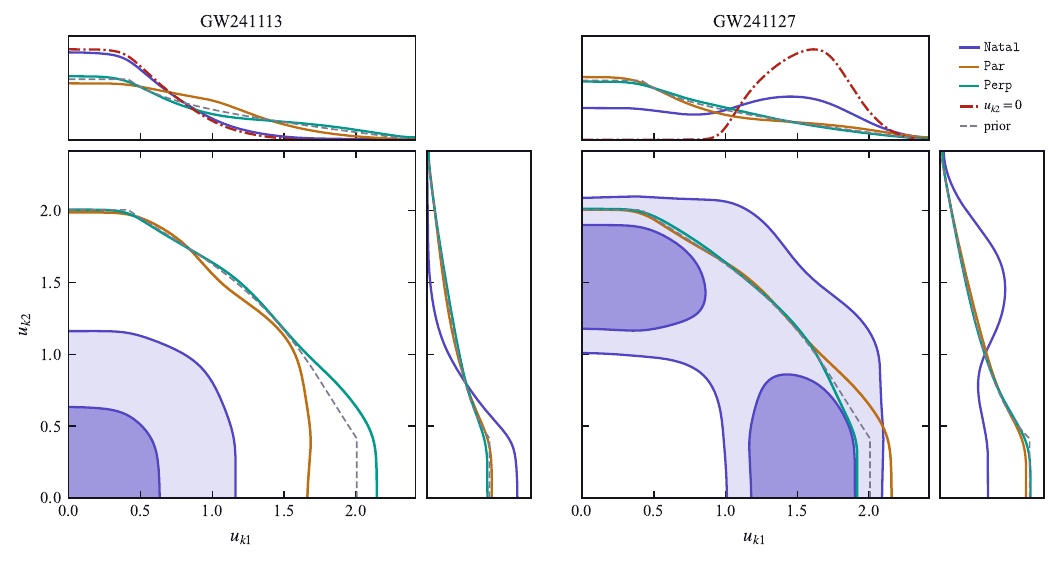}
    \caption{Joint constraints on the normalized kicks at CC1 and CC2 for GW241113\_163507 (left) and GW241127\_061008 (right). Purple shading and contours show the $50\%$ and $90\%$ credible regions under \Snat; orange and teal curves show the $90\%$ contours for \Spar and \Sperp. The dashed gray curves show the $90\%$ prior contour. The red dashed--dotted curve in the $u_{k1}$ marginal shows the restricted model with $u_{k2}=0$, in which only CC1 reorients the orbit.}
    \label{fig:two_kicks}
\end{figure*}

\textit{Allowing both collapse kicks to vary.} We also repeat the event-level inference with both $u_{k1}$ and $u_{k2}$ free for two representative BBHs. GW241113\_163507 provides the strongest two-kick support for \Snat, whereas GW241127\_061008 provides the weakest. Their joint posteriors are shown in Figure~\ref{fig:two_kicks}.

For GW241113\_163507, \Snat favors small kicks at both collapses and is mildly preferred to \Siso. For GW241127\_061008, the inherited-spin model instead requires substantial orbital reorientation distributed between the two collapses and is mildly disfavored. Neither kick is individually required to be large because their effects are degenerate. If $u_{k1}=0$ is imposed so that only CC2 tilts the orbit, GW241127\_061008 again requires a large kick. Allowing a nonzero second kick therefore weakens the attribution of the observed tilt specifically to CC1, but it does not remove the need for a large orbital reorientation under spin inheritance.

\textit{GW230529: a neutron star--BH candidate with a possibly retrograde primary spin.} GW230529 is most probably a neutron star--BH merger containing a lower-mass-gap primary of approximately $3.6\,\Msun$ and a secondary of approximately $1.4\,\Msun$, although the compact object identities cannot be determined conclusively from the GW data alone~\citep{LIGOScientific:2024elc}. The LVK analysis found $\chi_{1z}=-0.11^{+0.19}_{-0.35}$ and an $83\%$ posterior probability that $\chi_{1z}<0$. This inference is correlated with the mass ratio: more nearly equal component masses permit more negative values of the primary aligned-spin component.

\begin{figure}[t]
   \includegraphics[width=\columnwidth]{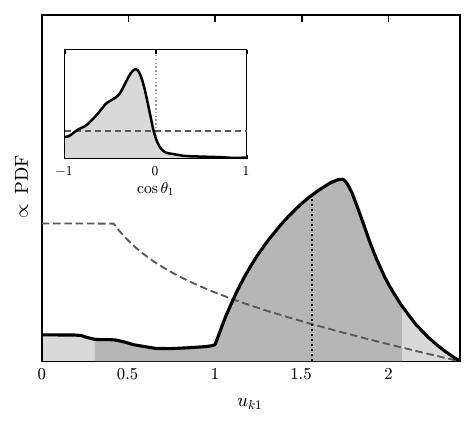}
    \caption{Natal model posterior for the normalized CC1 kick of GW230529 using the combined precessing PE analysis with a low secondary-spin prior. The light-gray fill shows the full posterior, the darker region marks its central $90\%$ credible interval,  and the dashed curve shows the survivability prior. The inset shows the corresponding PE posterior for $\cos\theta_1$; the dashed horizontal line shows the uniform PE prior. The primary tilt is $\cos\theta_1=-0.35^{+0.37}_{-0.53}$, while the kick posterior favors large values but remains broad.}
    \label{fig:gw230529}
\end{figure}

We analyze GW230529 separately from the BBH catalog using the combined precessing PE samples with a low secondary-spin prior, appropriate when the $\simeq1.4\,\Msun$ object is a neutron star. These samples give $\cos\theta_1=-0.35^{+0.37}_{-0.53}$, with $94\%$ of the posterior at $\cos\theta_1<0$. Under \Snat with $u_{k2}=0$, the kick posterior favors large values but remains broad and includes the low-kick regime, as shown in Figure~\ref{fig:gw230529}. The corresponding preference for \Siso over \Snat is modest. The \Spar and \Sperp results are indistinguishable from \Siso at this precision.

GW230529 is particularly informative because its component masses provide an independent reason to consider isolated binary evolution. Population synthesis calculations produce binaries with masses consistent with GW230529~\citep{Zhu:2024cvt, Chandra:2024ila}. By contrast, neutron star--black hole (NSBH) binaries are generally more difficult to assemble efficiently in many dynamical environments than BBHs, and some dynamical-capture calculations predict a smaller dynamical contribution to the NSBH merger rate than that from isolated binary evolution~\citep{Hoang:2020gsi}. These considerations do not exclude a dynamical origin, but they make isolated evolution a natural interpretation of GW230529.

That interpretation, however, creates a potential tension with the primary spin. Traditional population synthesis models can reproduce the component masses, but the commonly adopted inherited-spin prescription, \Snat, has difficulty producing a retrograde primary. Under \Snat, a retrograde tilt requires a natal kick comparable to or greater than the precollapse orbital velocity, which also disrupts a large fraction of binaries. The \Siso, \Spar, and \Sperp prescriptions instead provide concrete examples in which collapse-related processes generate a BH spin whose direction is not inherited from the progenitor rotation. They can therefore produce a retrograde spin component without requiring a large kick to reorient the orbital plane. GW230529 may point to alternative spin-generation physics within isolated binaries rather than necessarily to a different formation channel.

\subsection{Model systematics and limitations}
\label{sec:LVK_sys}

The analyses above isolate the dependence of the inferred natal kicks on the assumed spin-direction model by imposing several restrictive assumptions. We now discuss how these assumptions affect the interpretation of the results and how they can be tested with future observations.

\textit{Formation channel composition.} We have assumed that all 257 BBH mergers formed through isolated binary evolution and that a single spin-generation mechanism applies across the population. If part of the catalog instead formed through dynamical assembly in stellar clusters, galactic nuclei, or active galactic nuclei (AGN) disks, its spin-tilt distribution could differ from the four isolated-evolution prescriptions considered here. Dynamical populations are often approximated as isotropic, in which case they would contribute events resembling \Siso and reduce the fraction of large misalignments that \Snat must explain through natal kicks~\citep{Gerosa:2013laa, Rodriguez:2016vmx, Vitale:2015tea, Talbot:2017yur, Farr:2017uvj, Stevenson:2017dlk, Fishbach:2022lzq, Baibhav:2022qxm}.

Quantifying this possibility requires a hierarchical mixture analysis that assigns probabilistic formation channel membership rather than classifying individual events solely from their spin tilts.

The four models are also idealized in their treatment of spin generation. Apart from the tidal contribution considered in Section~\ref{sec:pop}, each BH is assigned a single source of spin. In reality, inherited core rotation, stochastic angular momentum transport, asymmetric collapse, accretion, and tides could contribute simultaneously, making the final BH spin the vector sum of components with different preferred directions. The superposition of these components could shift, broaden, or merge the two peaks characteristic of the pure \Spar and \Sperp models, potentially producing a single off-alignment peak similar to that discussed for the interacting populations in Section~\ref{sec:pop}. The relative importance of the different contributions could also vary with BH mass, metallicity, and binary separation. Our aim here is to isolate the observable signatures of the simplest limiting cases. A more comprehensive analysis should allow multiple contributions to the spin of each BH, together with their variation across the BBH population.

\textit{Second collapse and secondary spin.} Setting $u_{k2}=0$ neglects any orbital reorientation at CC2. As demonstrated by the two-kick analysis above, allowing both kicks to vary can distribute the required reorientation between the two collapses, weakening the attribution of the observed primary tilt specifically to CC1.

We also omit $\cos\theta_2$ from the tilt-dependent likelihood because it is weakly constrained for most individual events. However, several events have meaningfully constrained secondary tilts, and $\cos\theta_1$ and $\cos\theta_2$ can be correlated in the joint PE posterior. Replacing that joint information with the marginalized $\cos\theta_1$ posterior can therefore shift the inferred tilt distribution and introduce biases in both the kick inference and the model comparison. This simplification also reduces sensitivity to tidal alignment and other processes affecting the secondary spin between the two collapses~\citep{Bavera:2021evk}. A more complete analysis should retain the joint tilt posterior, particularly as the catalog and measurement precision improve.

\textit{First-born primary.} We identify the more massive component at merger with the first-born BH. Binary mass transfer can reverse the component-mass ordering, in which case the measured primary tilt would not directly trace the spin--orbit misalignment generated at CC1. A more complete analysis should marginalize over the possible correspondence between birth order and merger component labels.

\textit{Orbital properties.} We assume $\beta=1$ and circular precollapse orbits. Allowing mass loss or eccentricity would change the disruption boundaries and the accessible tilt ranges in Equations~(\ref{eq:cosLLRange})--(\ref{eq:psurv}), and could shift the inferred normalized kicks. The extent of these changes should be quantified in a future sensitivity analysis.

These assumptions define the scope of our conclusions. The event-level analysis provides a controlled comparison of spin-direction models and should not be interpreted as a direct measurement of the physical BH natal kick distribution.

Finally, we emphasize that only a small number of events in the catalog carry
enough spin information to say anything about their origin. The systems
highlighted here are those whose primary tilts are
measured well enough to prefer one spin-direction model over another, and even
for these the log Bayes factors remain of order unity. For the large majority
of the 257 BBHs, the primary-tilt posterior is broad, and the four models are
effectively indistinguishable. Spin tilts are among the least well-measured
parameters in GW observations, and the event-level analysis presented here
should therefore be read as a demonstration of how the assumed origin of the BH
spin changes the inferred kick, rather than as a classification of individual
systems.

The information that individual events lack is, however, recoverable in
aggregate. Small but correlated shifts in tilt across many weakly informative
events combine into a measurable population signature, so the features that
distinguish \Spar, \Sperp, \Siso, and \Snat---the peaks in
$\cos\theta_{1,2}$, the suppression of in-plane or aligned spins, and the
symmetry of \chieff about zero---are most naturally accessed at the
population level rather than event by event. Establishing which spin-generation
mechanism the data favor therefore requires a hierarchical population analysis
with selection effects, and most plausibly one that allows several formation
channels and several spin-generation mechanisms to operate simultaneously, with
their relative contributions inferred from the data rather than fixed in
advance. We leave such an analysis to future work.

\section{Summary and Conclusions}
\label{sec:conclusions}

BBHs detected by LVK are widely expected to be separable by their spin orientations, with dynamically assembled binaries isotropically oriented and field binaries aligned with the orbital angular momentum (Section~\ref{sec:intro}). For field binaries, this rests on the assumption that BH spins are inherited from the progenitor star. In this study, we explored spin-generation mechanisms that break that assumption. In Section~\ref{sec:LS}, we examine the distribution of spin--orbit tilts for four distinct models pertaining to the orientation of BH spins.  Each of these models has unique characteristics that depend on the natal kick magnitude, which are summarized in Table~\ref{tab:spin_summary}. In Section~\ref{sec:pop}, we use our models to make predictions about the distribution of spin--orbit tilts for a synthetic population. We find that the observed distribution is influenced by a complex interplay of various factors such as  kick magnitudes, evolutionary history, and tidal interactions. Distinct mechanisms that result in the spin-up of BHs will produce unique signatures in their spin orientations that can be detected in the population of merging BHs  observed by LVK. In Section~\ref{sec:LVK}, we compare these models with current LVK observations. Highly misaligned systems are often interpreted as evidence for dynamical formation because inherited-spin models struggle to produce them. We show that alternative spin-generation mechanisms can instead produce such systems naturally in isolated binaries, without requiring the entire misalignment to arise from kick-driven orbital reorientation. The observational implication is that neither precession nor $\chieff<0$ is, by itself, sufficient evidence for dynamical assembly.
For some of our models, we observe the appearance of two peaks in the spin--orbit tilt distribution. \Siso, \Sperp, and \Spar have symmetric distributions around $\cos\theta_{1,2}=0$, but this symmetry breaks when we allow tides to spin up the BHs along the orbital angular momentum. This can suppress the peak at $\cos\theta<0$ and amplify the peak at $\cos\theta>0$. The predicted tilt distributions are qualitatively compatible with the broad LVK reconstruction and with previously reported possible structure near $\theta\simeq60^\circ$~\citep{Vitale:2022dpa, Edelman:2022ydv, Callister:2023tgi}. This similarity supports the possibility that BH spin directions are influenced by the natal kick process, and these models offer a physical rationale for off-aligned structure that is difficult to produce under simple inherited-spin assumptions. Confirming this interpretation, however, will require a population-level analysis that allows multiple spin-generation mechanisms, such as those explored separately in Section~\ref{sec:pop}, to operate simultaneously and compares their combined predictions with future data. The location of the peaks depends on a complex combination of mass loss, kick magnitudes, pre-CC orbital separation, and tidal interactions; our assumptions regarding these quantities are oversimplified and may not fully capture the range of complexities encountered in binary evolution. Thus, a more detailed modeling approach is needed to make comparisons with observations.

In our assumptions, we consider that only one mechanism, in addition to tides, is capable of imparting spins. However, there is a possibility that multiple mechanisms may be at work in different regions of the parameter space. For instance, for low-mass BHs, large kicks could lead to spin-up according to either the \Spar or \Sperp model. On the other hand, for heavy BHs that undergo direct collapse without any asymmetric mass ejection and natal kicks, the infalling convective layers could result in spin-up according to the \Siso model (as suggested by \citet{Gilkis:2014rda, 2016ApJ...827...40G, Quataert:2018gnt, Antoni:2021xzs, Antoni:2023yxs}). This could result in distinct imprints on spin magnitudes and spin--orbit tilts for varying masses.

If BHs are spun up by kicks during formation, we assume in the \Spar and \Sperp models that their spins are proportional to the kicks. This leads us to expect a correlation between BH mass and spin magnitudes: under the reduced-kick prescription, the spin magnitude decreases with increasing BH mass. Similarly, effective spins have a decreasing width with increasing mass. Although no strong correlations have been found between BH mass and effective spin or spin magnitude, future GW observing runs may reveal more obscure features and correlations in even hard-to-measure parameters. It is also possible that only one of our assumptions about spin magnitude and spin--orbit tilt holds. For example, the spin-kick alignment model by \citet{Janka:2021deg} does not have spin proportional to kicks. In that case, despite the characteristic spin--orbit tilt distribution in the \Spar model, one might not find the correlation between mass and individual BH spins or effective spins. On the other hand, \citet{Muller:2018utr} found an isotropic distribution for spin--orbit tilts but spins proportional to kicks. In that case, even the isotropic spin model will have mass-spin correlations opposite to what was assumed here. While large kicks lead to more conspicuous spin--orbit tilt and spin magnitude distributions, they also tend to disrupt the binaries. Binaries that receive small kicks can survive, but the features in spin distribution are less noticeable. Moreover, the measurement of BH spins is a challenging task, which makes it difficult to study and resolve obscure features related to them.

The orientation of spin in BBHs not only provides insights into their origin but also helps understand the mechanisms that give BHs their spin. The way BHs acquire their spin affects the distribution of parameters like $\theta_{1,2}$, $\chi_{1,2}$, and $\chieff$. Analyzing these distributions can provide important information about the exact mechanisms behind BH spin generation. With the upcoming observing runs of LIGO and future detectors, we have a unique opportunity to solve the mysteries surrounding BH spins. By studying the imprints left by the process of BH spin generation, we can improve our understanding of the complex processes that govern their formation.

\section*{Acknowledgments}
V.B. acknowledges support from CIERA, Northwestern University, and the NASA Hubble Fellowship grant HST-HF2-51548.001-A awarded by the Space Telescope Science Institute, which is operated by the Association of Universities for Research in Astronomy, Inc., for NASA, under contract NAS5-26555. V.K. was supported by the Gordon and Betty Moore Foundation (grant awards GBMF8477 and GBMF12341), through a Guggenheim Fellowship, and the D.I. Linzer Distinguished University Professorship fund.
 This research was supported in part through the computational resources and staff contributions provided for the Quest high-performance computing facility at Northwestern University, which is jointly supported by the Office of the Provost, the Office for Research, and Northwestern University Information Technology.
This work used Delta at the National Center for Supercomputing Applications (NCSA) through allocation PHY260065 from the Advanced Cyberinfrastructure Coordination Ecosystem: Services \& Support (ACCESS) program, which is supported by US National Science Foundation grants
\#2138259, \#2138286, \#2138307, \#2137603, and \#2138296.

\bibliography{ref}{}
\bibliographystyle{aasjournal}

\end{document}